\documentclass[11pt, a4paper]{article}
\pdfoutput=1 
\usepackage[height=8.85in,width=6.75in]{geometry}
\usepackage{graphicx,rotating}     %usepackage without driver option
\usepackage[bookmarksopen,colorlinks=true,linkcolor=steelblue,
citecolor=orangered,urlcolor=darkred,linktoc=all]{hyperref}

\usepackage{amsmath,amssymb}
\usepackage{slashed}
\usepackage{bm}
\usepackage{bbm}
\usepackage{cite}
\usepackage{comment}
\usepackage[utf8]{inputenc}
\usepackage{accents}
\usepackage[footnotesize]{caption}
\usepackage{cancel}

\makeatletter
\@addtoreset{equation}{section}

\makeatother

\usepackage{xcolor}
\definecolor{darkred}{rgb}{0.7, 0., 0.}
\definecolor{orangered}{rgb}{1,0.27,0.}
\definecolor{steelblue}{rgb}{0.275,0.51, 0.706}
\definecolor{forestgreen}{rgb}{0.13,0.55,0.13}

\usepackage{tikz}
\usepackage{tikz-feynman}
\tikzfeynmanset{compat=1.0.0}
\pgfdeclarelayer{bg}    % declare background layer
\pgfsetlayers{bg,main}  % set the order of the layers (main is the standard layer)

\begin{document}
%%%%%%%%%%%%%%%%%%%%%%%%%%%%%%%%%%%%%%%

%%%%%%%%%%%%%%%%%%%%%%%%%%%%%%%%%%%%%%%
\hypersetup{pageanchor=false}
\begin{titlepage}

\begin{center}

\hfill UMN-TH-4325/24 \\
\hfill FTPI-MINN-24-16\\

\vskip 0.5in

{\huge \bfseries Momentum shift and on-shell recursion relation \vspace{0mm} \\ 
for electroweak theory
} \\
\vskip .8in

{\Large Yohei Ema,$^{1,2,a}$} \let\thefootnote\relax\footnote{$^a$ema00001@umn.edu}
{\Large Ting Gao,$^{1,b}$}\footnote{$^b$gao00212@umn.edu}
{\Large Wenqi Ke,$^{1,2,c}$}
\footnote{$^c$wke@umn.edu}
{\Large Zhen Liu,$^{1,d}$}
\footnote{$^d$zliuphys@umn.edu}
\vspace{1.5mm}
\\
{\Large Kun-Feng Lyu,$^{1,e}$}\footnote{$^e$lyu00145@umn.edu}
{\Large Ishmam Mahbub$^{1,f}$}
\footnote{$^f$mahbu008@umn.edu}
\vskip .3in
\begin{tabular}{ll}
$^{1}$ & \!\!\!\!\!\emph{School of Physics and Astronomy, University of Minnesota, Minneapolis, MN 55455, USA}\\
$^{2}$ & \!\!\!\!\!\emph{William I. Fine Theoretical Physics Institute, School of Physics and Astronomy,}\\[-.15em]
& \!\!\!\!\!\emph{University of Minnesota, Minneapolis, MN 55455, USA}\\
\end{tabular}

\end{center}
\vskip .6in

\begin{abstract}

 \noindent We study the All-Line Transverse (ALT) shift which we developed for on-shell recursion of amplitudes for particles of any mass. We discuss the validity of the shift for general theories of spin $\leq$ 1, and illustrate the connection between Ward identity and constructibility for massive spin-1 amplitude under the ALT shift. We apply the shift to the electroweak theory, and various four-point scattering amplitudes among electroweak gauge bosons and fermions are constructed. We show explicitly that the four-point gauge boson contact terms in massive electroweak theory automatically arise after recursive construction, independent of UV completion, and they automatically cancel the terms growing as (energy)$^4$ at high energy. We explore UV completion of the electroweak theory that cancels the remaining (energy)$^2$ terms 
and impose unitarity requirements to constrain additional couplings. 
The ALT shift framework allows consistent treatment in dealing with contact term ambiguities for renormalizable massive and massless theories, which we show can be useful in studying real-world amplitudes with massive spinors.
\end{abstract}

\end{titlepage}
%%%%%%%%%%%%%%%%%%%%%%%%%%%%%%%%%%%%%%%

%%%%%%%%%%%%%%%%%%%%%%%%%%%%%%%%%%%%%%%
\tableofcontents
\renewcommand{\thepage}{\arabic{page}}
\renewcommand{\thefootnote}{$\natural$\arabic{footnote}}
\setcounter{footnote}{0}
%\newpage
\hypersetup{pageanchor=true}
%%%%%%%%%%%%%%%%%%%%%%%%%%%%%%%%%%%%%%%

%%%%%%%%%%%%%%%%%%%%%%%%%%%%%%%%%%%%%%%
\section{Introduction}
\label{sec:introduction}
%%%%%%%%%%%%%%%%%%%%%%%%%%%%%%%%%%%%%%%

%%%%%%%%%%%%%%%%%%%%%%%%%%
\begin{figure}[t!] 
  \centering
    \includegraphics[width=0.9\textwidth]{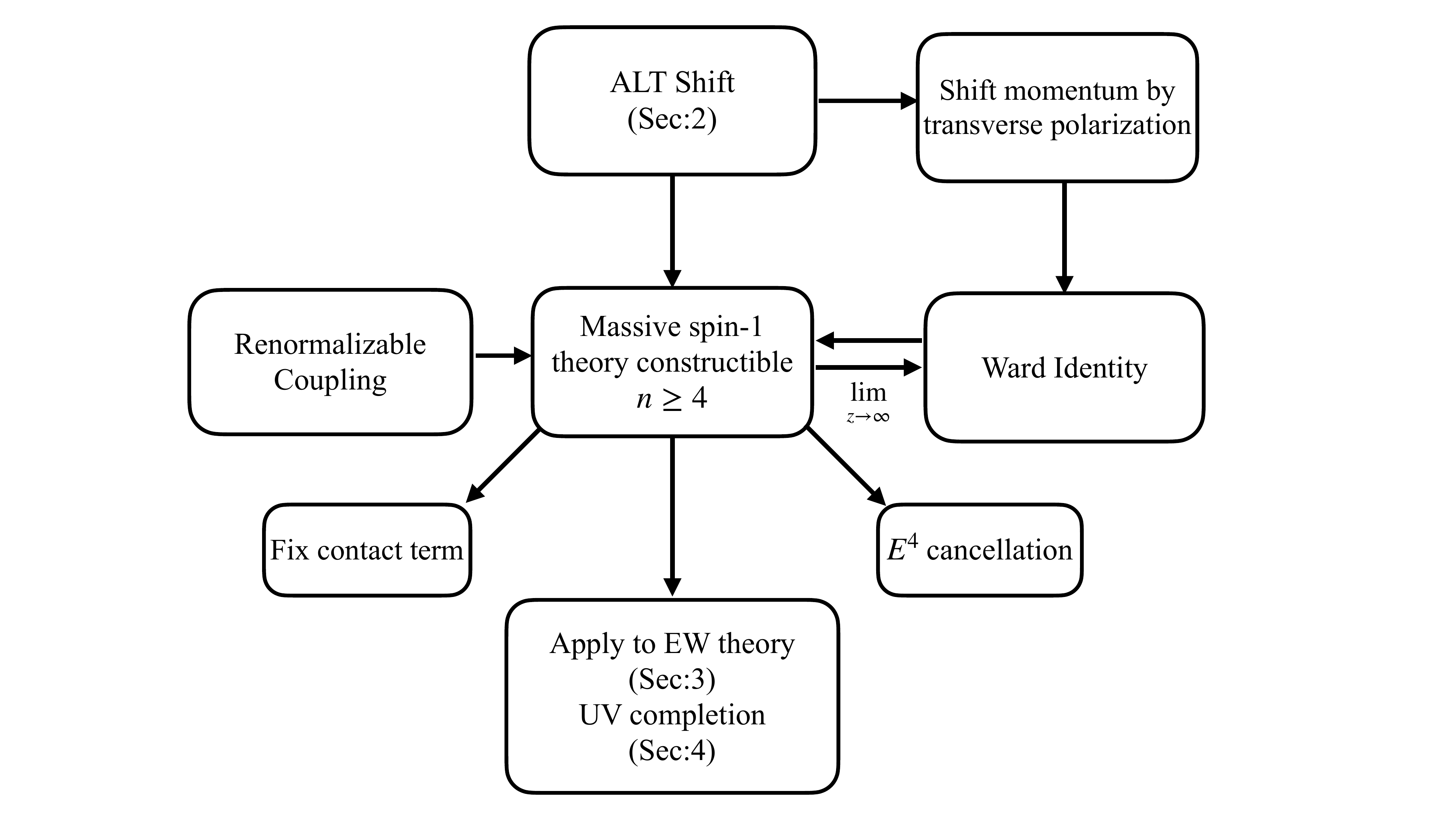}
\caption{Summary of recursion relation for renormalizable massive vector theory. 
}
\label{fig:schematic_flow_1}
\end{figure}
%%%%%%%%%%%%%%%%%%%%%%%%%%%%%%%
Modern methods of scattering amplitude exploit unitarity, locality and causality to construct tree-level amplitudes and explores quantum field theory. 
The on-shell recursion method developed in~\cite{Britto:2004ap,Britto:2005fq} has allowed the construction of higher-point amplitudes from on-shell lower-point amplitudes.
These methods were first applied to pure Yang-Mills theory and gravity~\cite{Bedford:2005yy,Cachazo:2005ca},
and subsequently extended to various theories (see e.g.~\cite{Arkani-Hamed:2008owk,Cohen:2010mi,Elvang:2013cua,Henn:2014yza,Cheung:2015ota,Cheung:2017pzi,Travaglini:2022uwo} for reviews). 
Although on-shell recursion of massless external particles as well as its extension to a mixture of massive and massless particles has been studied~\cite{Badger:2005zh,Badger:2005jv,Ozeren:2006ft,Boels:2007pj,Ballav:2020ese}, recursion techniques for general theories with all massive particles are still part of ongoing research~\cite{Cohen:2010mi,Arkani-Hamed:2017jhn, Herderschee:2019dmc, Herderschee:2019ofc, Franken:2019wqr,Wu:2021nmq,Ema:2024vww}.

The on-shell method is based on analytical continuation of an amplitude by shifting external momenta 
with a complex variable $z$. Using Cauchy's theorem, the amplitude is written as a sum of a contribution from 
poles at finite $z$ where intermediate particles go on-shell and the amplitude factorizes 
into products of lower-point amplitudes, 
and a non-factorizable contribution at $z \rightarrow \infty$. 
An amplitude is ``on-shell constructible'' if there exists a momentum shift under which the latter contribution is absent. The contact term ambiguity arises here as one might find the constructed amplitudes differ from the correct prediction of the underlying theory (e.g., from Feynman diagram approach for a given theory) by contact terms, which can come from either new theory inputs or inconsistent momentum shift. A useful momentum shift for massive amplitudes would allow us to resolve these contact term ambiguities, for instance, as recently demonstrated in Ref.~\cite{Ema:2024vww}. 
In massless theories, conditions to satisfy the on-shell constructibility are well understood, starting with the famous example of the BCFW-shift~\cite{Britto:2004ap,Britto:2005fq}.
On the contrary, momentum shifts are less explored in massive theories,
and this is the direction we pursue here.

In this paper, we apply the All-Line Transverse (ALT) shift we proposed in~\cite{Ema:2024vww},
which deforms external momenta using transverse polarization vectors,
to amplitudes involving massive $W$- and $Z$-bosons in the electroweak (EW) theory, as schematically shown in \autoref{fig:schematic_flow_1}. With the dimensional analysis and under the assumption that the theory satisfies the Ward identity in the massless limit, 
we show that four- and higher-point amplitudes
are constructible under the ALT shift in a general renormalizable theory.\footnote{
	with the exception of four scalar amplitudes,
	which contain a non-factorizable term given by $A_4 = \lambda$ with a coupling constant $\lambda$.
	Unless imposing additional symmetries such as supersymmetry,
	this amplitude is not constructible by any momentum shift.
}
In particular, we explicitly construct $W$- and $Z$-boson four-point amplitudes, including the contact terms, solely from the three-point amplitudes.
This requires an introduction of only two dimensionless couplings, $g_{WWZ}$ and $g_{WW\gamma}$, for three-point amplitudes among $W$ and $Z$/photon. 
The contact terms arise from the ALT shift-dependent parts of the three-point amplitudes,
and therefore this would be achieved only with an explicit introduction of the momentum shift.
This is in contrast to the approaches in~\cite{Bachu:2019ehv,Liu:2022alx,Christensen:2024xzs,Christensen:2024bdt}
where the amplitudes are glued without momentum shift, which in general raises the contact term ambiguity (see e.g.~\cite{Christensen:2022nja,Lai:2023upa,Ema:2024vww} in the context of QED with massive fermions).
The contact term derived by the ALT shift automatically guarantees the cancellation of the term
growing as (energy)$^4$ at high energy 
in all-longitudinal $W$-/$Z$-scatterings, and this is independent of UV completion of the EW theory.

We then move our focus to different UV completions of the EW theory.
Even though the (energy)$^4$ terms are canceled, the amplitudes still grow as (energy)$^2$ without additional input, 
and we require the cancellation of these terms by adding weakly coupled particles. With the assumption of adding only one neutral scalar, we show that it naturally fits into a Higgs-doublet model. We then consider adding one neutral and one doubly charged scalar, which naturally leads to a Higgs-triplet model, and briefly discuss UV completion by adding vector particles.

This paper is organized as follows. In Sec.~\ref{sec:mom_shift},
the general formalism of the on-shell recursion relation is reviewed. 
We then introduce the ALT shift and discuss its large-$z$ behavior that determines the on-shell constructibility. We highlight an important feature of ALT shift: that the large-$z$ limits of the shifted amplitudes take a form that Ward Identity can be directly imposed. In Sec.~\ref{sec:EW}, we apply the shift for on-shell construction of $W$- and $Z$-boson amplitudes. 
Different UV completions are discussed in Sec.~\ref{sec:UV}.
Finally, Sec.~\ref{sec:conclusion} is devoted to a summary of our results and possible outlooks.
The appendix consists of three parts. In App.~\ref{app:convention}, we summarize the massive spinor formalism 
used throughout this paper and define our notations and conventions.
In the main text, we express our calculations in the little group covariant massive spinor representation~\cite{Arkani-Hamed:2017jhn}.
For readers who are more used to the vector notation, we express our results of the EW theory and its UV completions
in the vector representation in Apps.~\ref{app:EW_vector} and~\ref{app:UV_vector}, respectively.

%%%%%%%%%%%%%%%%%%%%%%%%%%%%%%%%%%%%%%%
\section{Momentum shift and massive constructive method}
\label{sec:mom_shift}
%%%%%%%%%%%%%%%%%%%%%%%%%%%%%%%%%%%%%%%

In this section, we begin with a brief overview of the on-shell methods for tree-level amplitudes (see e.g.~\cite{Elvang:2013cua,Henn:2014yza,Cheung:2017pzi,Travaglini:2022uwo} for reviews), 
and introduce the All-Line Transverse (ALT) shift. This method will be used in Secs.~\ref{sec:EW} and~\ref{sec:UV} to construct tree-level amplitudes in the EW theory.

%%%%%%%%%%%%%%%%%%%%%%%%%%%%%%%%%%%%%%%
\subsection{Recursion relation and factorization}
%%%%%%%%%%%%%%%%%%%%%%%%%%%%%%%%%%%%%%%
\label{sec:rec_and_fact}
To construct higher-point amplitudes from lower-point on-shell amplitudes, we shift the momenta of external particles as
\begin{align}
	p_i \to \hat{p}_i = p_i + z q_i,
	\quad
	\sum_{i} q_i = 0,
	\label{eq:mom_shift}
\end{align}
where $i$ runs over $1$ to $n$ with $n$ the total number of the external particles, $q_i$ is an arbitrary momentum
which we later choose to be proportional to the transverse polarization vector, 
and the latter condition guarantees the total momentum conservation.
We require
\begin{align}
	p_i \cdot q_i = q_i^2 = 0,
\end{align}
so that the on-shell condition for external particles is not modified, $\hat{p}_i^2 = p_i^2 = m_i^2$,
where $m_i$ is the mass of the particle $i$. The momentum shift promotes the amplitude to a complex function, $\hat{A}_n(z)$,
and the original amplitude is given by $A_n = \hat{A}_n(z=0)$. Focusing on tree-level amplitudes, $\hat{A}_n(z)$ has only poles (and no branch cuts) in the complex $z$-plane. Each pole corresponds to an intermediate particle being on-shell, where the amplitude factorizes into
a product of lower-point on-shell amplitudes.
Therefore, by using the complex analysis, we can express the amplitude as
\begin{align}
	 {A}_n &= \frac{1}{2\pi i}\oint_{z=0} \frac{dz}{z} \hat{A}_n(z)
	=  -\sum_{z = z_I}\sum_\lambda \mathrm{Res}_{z=z_I}
	\left[\hat{A}^{(\lambda)}_{n-m+2}\frac{1}{z}\frac{1}{\hat{p}_I^2 - m_I^2}
	\hat{A}^{(\bar{\lambda})}_{m}\right]
	+ B_\infty,
\end{align}
where $z_I$ satisfies $\hat{p}_I^2 - m_I^2 = 0$ with $m_I$ the mass of the intermediate particle and $\hat{p}_I$
the partial sum of the external momenta.  
$\lambda$ denotes the helicity of the intermediate particle with $\bar{\lambda}$ its little-group conjugate (we will explain it later), and  $B_\infty$ is the boundary contribution at $\vert z\vert \to \infty$.
In our case, the pole condition is solved as
\begin{align}
	\hat{p}_I^2 - m_I^2 = 0 ~~\Longrightarrow~~
	 z^{\pm}_I  =  \dfrac{1}{q_I^2}\left[
	-p_I \cdot q_I \pm \sqrt{(p_I \cdot q_I)^2 - (p_I^2 - m_I^2)q_I^2}\,\right].
\end{align}
The ALT shift (which we specify below) satisfies $q_i \cdot q_j \neq 0 $ for $i\neq j$, resulting in $q_I^2 \neq 0$, and therefore we have a pair of the poles, $z_I^\pm$, for each intermediate particle.
By using
\begin{align}
	\frac{1}{\hat{p}_I^2 - m_I^2} = \frac{1}{p_I^2 - m_I^2}\frac{z_I^+ z_I^-}{(z-z_I^+)(z-z_I^-)},
\end{align}
the amplitude is then written as
\begin{align}
	 {A}_n &=   
	\sum_{I} 
	\frac{1}{p_I^2-m_I^2}
	\frac{1}{z_I^+ - z_I^-}
	\sum_{\lambda}
	\left[z_I^+\hat{A}^{(\lambda)}_{n-m+2}(z_I^-)\times 
	\hat{A}^{(\bar{\lambda})}_{m}(z_I^-)
	- (z_I^+ \leftrightarrow z_I^-) \right]
	+ B_\infty.
	\label{eq:recursion_general}
\end{align}
This tells us that if $B_\infty = 0$, or if $\lim_{z\to \infty} \hat{A}_n(z) = 0$,
the higher-point amplitudes are expressed by lower-point on-shell amplitudes,
implying the on-shell constructibility of the amplitudes. Note that such on-shell constructibility does not mean there exist no contact term in the theory, but rather these contact terms are dependent on the lower point vertices through the consistent amplitude construction. Later in this section, we investigate the large-$z$ behavior of the shifted amplitude $\hat{A}_n(z)$ that controls $B_\infty$ under our proposed ALT shift.

We also remark that for a given massive theory, it is possible that amplitudes with particular combinations of spin projections are on-shell constructible, and not all possible spin projections are on-shell constructible for a given shift. However, since amplitudes are little group covariant, we can act the spin-raising and spin-lowering operators on the amplitude with one particular spin-projection to construct the rest of the amplitudes. These operators are generators of the little group SU(2), and they are defined in the helicity basis as
\begin{align}
	     J_i^- = -\lvert i \rangle_\alpha \dfrac{\partial}{\partial \lvert \eta_i \rangle_\alpha} 
	     - [ \eta_i \lvert_{\dot{\alpha}} \dfrac{\partial}{\partial [ i\lvert_{\dot{\alpha}}}, 
	     \qquad
	     J_i^+ = \lvert \eta_i \rangle_\alpha \dfrac{\partial}{\partial \lvert i \rangle_\alpha} 
	     + [ i \lvert_{\dot{\alpha}} \dfrac{\partial}{\partial [ \eta_i\lvert_{\dot{\alpha}}}.
\end{align}
In calculations below, we will show this operation leads to consistent and correct amplitudes for all longitudinal scatterings, which are not directly constructible under ALT shift.

%%%%%%%%%%%%%%%%%%%%%%%%%%%%%%%%%%%%%%%
\subsection{All-Line Transverse (ALT) shift}
\label{subsec:ALTshift}
%%%%%%%%%%%%%%%%%%%%%%%%%%%%%%%%%%%%%%%

In the rest of the paper, we focus on a momentum shift scheme which is universally applicable for both massive and massless amplitudes, dubbed ALT shift, introduced in Ref.~\cite{Ema:2024vww}. The ALT shift is based on the helicity decomposition of the spinors, given by
\begin{align}
	(p_i)_{a\dot{a}} = \vert i\rangle_a [i\vert_{\dot{a}} - \vert\eta_i\rangle_a [\eta_i\vert_{\dot{a}}.
\end{align}
For spinor notation and convention, see App.~\ref{app:convention}. We construct the ALT shift after declaring the helicity of all external particles, and the spinor shifts are defined such that external Dirac spinors and polarization vectors are unshifted. For spin-1/2 particles and the transverse modes of spin-1 particles, we shift the spinors such that
\begin{align}
	\begin{cases}
	\vert i \rangle \to \vert \hat{i}\rangle = \vert i\rangle + z c_i \vert \eta_i\rangle
	& \mathrm{for}~~\lambda_i = +, \vspace{1mm} \\
	\vert i ] \to \vert \hat{i}] = \vert i] + z c_i \vert \eta_i]
	& \mathrm{for}~~\lambda_i = -,
	\end{cases}
      \quad \mathrm{or}~~~
	\begin{cases}
	\vert \eta_i ] \to \vert \hat{\eta}_i] = \vert \eta_i] + z c_i \vert i]
	& \mathrm{for}~~\lambda_i = +, \vspace{1mm} \\
	\vert \eta_i \rangle \to \vert \hat{\eta}_i\rangle = \vert \eta_i\rangle + z c_i \vert i\rangle
	& \mathrm{for}~~\lambda_i = -,
	\end{cases}	
 \label{transverseshifts}
\end{align}
and the other spinors are unshifted.
Here, $\lambda_i$ is the helicity index of the particle $i$
and $c_i$ is a numerical constant required to satisfy the total momentum conservation. We can extend the shift to incorporate massless particles with the understanding that the spinors $\vert \eta_i \rangle, \vert \eta_i]$ are now replaced by arbitrary reference spinors, usually denoted as $\vert \xi_i\rangle, \vert \xi_i]$, for these particles.
For the longitudinal ($L$) modes of spin-1 particles, there are two choices:
\begin{align}
	\begin{cases} 
	\vert i\rangle \to \vert \hat{i}\rangle = \vert i\rangle + z\dfrac{c_i}{2}\vert \eta_i \rangle, \vspace{1mm}\\
	[\eta_i\vert \to [\hat{\eta}_i\vert =  [\eta_i\vert - z\dfrac{c_i}{2}[i\vert,
	\end{cases}
	\mathrm{or}~~~
	\begin{cases}
	[i\vert \to [ \hat{i}\vert = [ i\vert + z\dfrac{c_i}{2}[ \eta_i \vert, \vspace{1mm}\\
	\vert\eta_i\rangle \to \vert\hat{\eta}_i\rangle = \vert\eta_i\rangle - z\dfrac{c_i}{2}\vert i\rangle,
	\end{cases}
	\mathrm{for}~~\lambda_i = L,
	\label{eq:all_trans_long}
\end{align}
and both choices equally work for our purpose. The Dirac spinors are given by 
\begin{align}
	u^{(+)} = \begin{pmatrix} \vert \eta_i\rangle_a \\ \vert i]^{\dot{a}}\end{pmatrix},
	\quad
	u^{(-)} = \begin{pmatrix} \vert i\rangle_a \\ \vert \eta_i]^{\dot{a}}\end{pmatrix},
	\quad
	v^{(+)} = \begin{pmatrix} \vert \eta_i\rangle_a \\ -\vert i]^{\dot{a}}\end{pmatrix},
	\quad
	v^{(-)} = \begin{pmatrix} \vert i\rangle_a \\ -\vert \eta_i]^{\dot{a}}\end{pmatrix},
\end{align}
while the polarization vectors for the gauge boson are expressed as
\begin{align}
	\epsilon^{(+)} = \sqrt{2}\frac{\vert \eta_i\rangle [ i \vert}{m_i},
	\quad
	\epsilon^{(-)} = -\sqrt{2}\frac{\vert i\rangle [ \eta_i \vert}{m_i},
	\quad
	\epsilon^{(L)} = \frac{\vert i\rangle [i \vert + \vert \eta_i\rangle [\eta_i \vert}{m_i}.
\end{align}
Hence the external polarization vectors and Dirac spinors are invariant under the ALT shift. With the above shift of the spinors, we see that the momenta are shifted by the transverse polarization vector
\begin{align}
	p_i \to p_i + z q_i,~~q_i = c_i m_i\epsilon^{(\pm)}(p_i).
\end{align}
For the external gauge boson with transverse helicity, 
the shifted momentum is equal to the same helicity transverse polarization vector, and similarly for the fermion.
For the longitudinal gauge boson, we can shift the momentum by either positive or negative transverse polarization vectors, corresponding to the two choices in Eq.~\eqref{eq:all_trans_long}.\footnote{ We do not shift the momentum by the longitudinal polarization vector since this choice shifts the external longitudinal polarization vector and modifies the on-shell condition.} The transverse polarization vector does not have the temporal component in this basis, and hence we can satisfy the total momentum conservation, i.e.~the second equation in Eq.~\eqref{eq:mom_shift}, with non-zero $c_i$ for four and higher-point amplitudes.

%%%%%%%%%%
\subsubsection*{Large-$z$ behavior}
%%%%%%%%%%

As discussed in~\cite{Ema:2024vww}, a simple dimensional analysis allows us to estimate the large-$z$
behavior of the amplitude under the ALT shift. To see this, it is convenient to decompose the amplitude as~\cite{Cheung:2015cba,Ema:2024vww}
\begin{align}
	A_n = \left[\sum_{\mathrm{diagrams}}g \times F\right]
	\times \prod \epsilon \times \prod u,
\end{align}
where $g$ is the collection of the couplings, and $F$ is the amplitude with the external polarization vectors and fermion wavefunctions, which we denote collectively by $\prod \epsilon$ and $\prod u$, stripped off. ALT shift satisfies $q_i \cdot q_j \neq 0$ for $i\neq j$, the denominators of the propagators all scale as $z^2$. Combined with the numerator, we expect that $F$ scales at most as $\hat{F} \sim z^{\gamma}$ with $\gamma \leq [F]$ at large $z$, where $[F]$ is the mass dimension of $F$.\footnote{
We note here a subtle difference to our traditional mass/energy dimensional analysis. In our explicit calculation below, we encounter the structure $1/m_W^2$ arising from the completeness
relation of the intermediate polarization vectors. This could possibly increase the $z$ scaling counting by two each time. We assume that this factor is compensated by massive parameters
in the numerator \textit{without} increasing $z$ scaling, either due to the Ward identity or due to the couplings with positive mass dimensions. In other words, we focus on the theory that has a smooth massless limit, such as the gauge theory that is only spontaneously broken. Under this assumption, this factor does not modify our argument based on the dimensional analysis. 
}
The mass dimension of an $n$-point amplitude is $4-n$ which fixes $[F]$ as $[F] = 4-n - [g] -N_F/2$, where $N_F$ is the number of the external fermions and $[g]$ is the mass dimension of $g$. Since the ALT shift does not deform the external polarization vectors and fermion wavefunctions, we obtain the large-$z$ behavior of the amplitude as
\begin{align}
	\hat{A}_n \sim z^{\gamma},
	\quad
	\gamma \leq  4-n - [g] - \frac{N_F}{2}.
	\label{eq:large_z_naive}
\end{align}
If $\gamma < 0$, $B_\infty = 0$ and
the theory is on-shell constructible. The dimensional analysis suffices to prove the constructibility of four-point amplitudes with at least one pair of fermions as well as five- and higher-point amplitudes for renormalizable theories (which have $[g] \geq 0$).

%%%%%%%%%%
\subsubsection*{Ward identity and on-shell constructibility}
%%%%%%%%%%

The estimate~\eqref{eq:large_z_naive} based solely on the dimensional analysis
does not indicate the on-shell constructibility of four vector boson scatterings in a renormalizable theory. But if we have (at least) one transverse gauge boson in the external state,
large-$z$ behavior under ALT shift is 
improved by the Ward identity for massless gauge bosons. This is a feature of the ALT shift. Since we shift our momentum by the transverse polarization vector, 
the external momentum and polarization vector for the transverse mode become proportional to each other
at the leading order in the large $z$ limit,
\begin{align}
	\lim_{z\to \infty} \hat{p}_i \propto \epsilon_i^{(\lambda_i)},
\end{align}
and this allows us to use the Ward identity.
Saturating the inequality~\eqref{eq:large_z_naive} 
means that we consume all the dimensionful quantities by the shifted momentum $q_i~(\propto \epsilon_i^{(\pm)})$. To be concrete, let us take the particle $1$ as the transverse gauge boson, and extract the polarization vector of the particle $1$.
The leading order term saturating Eq.~\eqref{eq:large_z_naive} at large $z$
is then schematically expressed as
\begin{align}
	\lim_{z \rightarrow \infty}\hat{A}_n 
	&\sim g\,z^{4-n-[g] - N_F/2}  \big[\epsilon_{1}^{(\lambda_1)} \cdot F(q_1, q_2, ..., q_n)\big]
	\propto g\,z^{4-n-[g] - N_F/2} \big[ q_{1} \cdot F(q_1, q_2, ..., q_n)\big],
\end{align}
where $F$ depends only on the shifted momentum $q_i$ and it does not contain any other mass dimensionful parameters (such as $m_W$) except possibly for their ratios. Notice that we can think of the polarization vector of the particle $i$ as defined with respect to $q_i$, instead of $p_i$, since it satisfies $\epsilon_i \cdot q_i = 0$ in the ALT shift. This is a consequence of our choice of the momentum shift that does not shift the external polarization vectors. We can then \emph{impose} the Ward identity of the massless theory, resulting in
\begin{align}
	q_{1} \cdot F(q_1, q_2, ..., q_n) = 0,
\end{align}
meaning that the leading order term in large $z$ vanishes. For the amplitude with only the longitudinal modes, the above argument does not apply since $\epsilon_1^{(L)}$ is not proportional to the shifted momentum. Therefore, we have
\begin{align}
	\gamma \leq 4-n - [g] - \frac{N_F}{2} - c_T,
	\label{eq:large_z_Ward}
\end{align}
where $c_T = 1$ if there is at least one external transverse mode, and $c_T = 0$ if all the external gauge bosons are longitudinal. This holds as long as we require the Ward identity of the theory in the massless limit. This is the on-shell equivalent of declaring that we start from the gauge theory, which is broken only spontaneously.  

The relation~\eqref{eq:large_z_Ward} also indicates that we cannot construct the scalar four-point amplitude, but this is expected. For the scalar four-point amplitude, there exists no momentum shift that allows the on-shell construction
since we can always add the scalar four-point interaction, $\lambda \phi^4$,
in a renormalizable theory (in the Lagrangian formulation). This interaction is not related to any lower-point amplitudes, and hence they contribute to the boundary term (unless we assume e.g.~supersymmetry).

We have so far proved that renormalizable massive spin-1 theories satisfying Ward identity are on-shell constructible for $n \geq 4$ using ALT shift. In the previous subsection, we also showed the constructibility of a generic renormalizable massive spin-1 theory starts at $n \geq 5$, which indicates that the four-point contact term between vectors and scalars can be \textit{independent} in general. Now one can ask for the ALT shift, what space of massive spin-1 theories are on-shell constructible using only three-point amplitudes, and whether theories satisfying Ward identity are the only ones fulfilling such constructibility constraints. In this regard, for instance, we observed a connection between the two in the construction of $WWhh$ amplitude in Sec.~\ref{sec:WWhh_triplet}. The amplitude produces boundary contributions due to the presence of the contact term. Demanding on-shell constructibility using ALT shift, we can determine the relation among three-point couplings. We can derive the same relation if we instead require the amplitude to obey Ward identity in the massless limit, which illuminates the connection between on-shell constructibility for $n\geq 4$ and Ward identity. 

In the rest of this paper, 
we focus on the EW theory without additional higher dimensional operators and therefore assume $[g] \geq 0$.\footnote{We see ALT shift allows for a systematic discussion on the contract terms, extracting which part of the coefficient is determined from the consistent conditions from lower point amplitudes and which part is new input.}
The relation~\eqref{eq:large_z_Ward} is then sufficient for our purpose.
Higher dimensional operators, in general, tend to spoil on-shell constructibility 
since it is not necessarily related to lower-point amplitudes. Indeed, higher dimensional operators correspond to $[g] < 0$, and this makes the large-$z$ behavior under the ALT shift worse. In this regard, we may note that in the above discussion, we have focused on a transverse polarization of a single particle,
but one may ask if having more than one transverse modes further improve the boundary behavior with ALT shift.
If such an improvement exists, one may apply it for recursion relations in the presence of higher dimensional operators
to overcome the worse large-$z$ behavior.\footnote{
For recent developments in applying the massive spinor formalism to SM effective field theories, 
see~\cite{Shadmi:2018xan,Aoude:2019tzn,Durieux:2019eor,Durieux:2020gip,Balkin:2021dko,Liu:2023jbq}. 
}

%%%%%%%%%%%%%%%%%%%%%%%%%%%%%%%%%%%%%%%
\section{On-shell construction of electroweak theory}
\label{sec:EW}
%%%%%%%%%%%%%%%%%%%%%%%%%%%%%%%%%%%%%%%

We now construct the four-point amplitudes of the $W$- and $Z$-bosons in the EW theory using on-shell recursion relation.
In this section, we do not include the Standard Model (SM) Higgs,
assuming that only the EW gauge bosons and SM fermions exist in the spectrum.
We will see that the ALT shift correctly constructs the amplitudes, including the contact terms,\footnote{
This corresponds to the gauge four-point interactions in the Feynman diagrammatic language.
} solely from the three-point amplitudes,
independently of the presence/absence of the Higgs.\footnote{ Indeed, the constructibility is still valid because the Ward identity for massless gauge boson is independent of the details of Higgsing.}
The momentum-shifted parts of the three-point amplitudes play an essential role
in constructing the contact term as we see below, and this would not be achieved without introducing an explicit momentum shift.

Our result, with the contact term derived by the ALT shift, automatically guarantees 
the cancellation of the $\mathcal{O}(E^4)$ term of all-longitudinal gauge boson scatterings at high energy, where $E$ denotes the center-of-mass energy.
We still have finite $\mathcal{O}(E^2)$ terms in our amplitudes, and we have different possibilities of canceling this divergence, e.g.~by introducing the SM Higgs, as we discuss in Sec.~\ref{sec:UV}.
Thus, our calculation based on the ALT shift clarifies the different natures of the $\mathcal{O}(E^4)$ and $\mathcal{O}(E^2)$ terms in the four longitudinal gauge boson scatterings; the $\mathcal{O}(E^4)$ term is controlled by the gauge symmetry and the Ward identity, while the $\mathcal{O}(E^2)$ term requires additional input for the cancellation. This is, of course, well-known in the context of the Feynman diagrammatic calculations~\cite{LlewellynSmith:1973yud,Lee:1977yc,Lee:1977eg}, but we believe that we have derived this fact,  for the first time, based solely on the on-shell method.
This is in contrast to, e.g., the discussions in~\cite{Bachu:2019ehv,Liu:2022alx} 
where the authors do not introduce a momentum shift, and both the four-point contact terms and the Higgs contributions
are added by hand, on equal footing, to improve the high-energy behavior.

%%%%%%%%%%%%%%%%%%%%%%%%%%%%%%%%%%%%%%%%%
\subsection{Three-point amplitude}
\label{subsec:three-point_EW}
%%%%%%%%%%%%%%%%%%%%%%%%%%%%%%%%%%%%%%%%%

For the on-shell construction of the four-point EW amplitudes, we need three-point amplitudes involving 
massive spin-1 particles $W^\pm,Z$ with masses $m_W,m_Z$, and a massless spin-1 particle $\gamma$. We denote $W^+$ and $W^-$ as ``$1$" and ``$2$", and $Z$ or $\gamma$ as ``$3$". As discussed in~\cite{Arkani-Hamed:2017jhn,Bonifacio:2018vzv,Durieux:2019eor}, the minimal three-point amplitudes are\footnote{Here we comment on one observation of the minimal form of the three-point amplitude and its relation to constructibility. For amplitudes involving the exchange of massive bosons, the propagator has $pp/m^2$ in the numerator, and the amplitude takes the form $J \Pi J$ where $\Pi$ is the propagator and $J$ is the current associated to the gauge boson.
If we require gauge invariance in the massless limit (equivalently, Ward identity for massless gauge bosons), i.e.~$p^\mu J_\mu=\mathcal{O}(m) $, the $pp/m^2$ term in the propagator contracts with $J$ on both sides to give $\mathcal{O}(m)$. Without the minimal coupling, the $pp/m^2$ term will stay which in general spoils large-$z$ behavior, hence spoils constructibility.}
%%%%%%%%%%%%%%%%%%%%%%%%%%%%%%%%
\begin{align}
	A_{WWZ}^{(\lambda_1\lambda_2\lambda_3)}
	&= 
	\begin{tikzpicture}[baseline=(v)]
	\begin{feynman}[inline = (base.a), horizontal=p1 to v]
		\vertex (p1) [label=\({\scriptstyle W^+}\)];
		\vertex [below right = of p1] (v);
		\vertex [below left = of v,label=270:\({\scriptstyle W^-}\)] (p2);
		\vertex [right = of v, label=\({\scriptstyle Z}\)] (p3);
		\node [left = -0.25cm of v, blob, fill=gray] (vb);
		\begin{pgfonlayer}{bg}
		\diagram*{
		(p1) -- [photon,momentum=\({\scriptstyle p_1}\)] (v) -- [photon,rmomentum'=\({\scriptstyle p_2}\)] (p2),
		(v) -- [photon, rmomentum'=\({\scriptstyle p_3}\)] (p3),
		};
		\end{pgfonlayer}{bg}
	\end{feynman}
	\end{tikzpicture}
	= \frac{g_{WWZ}}{\sqrt{2}m_W^2 m_Z} \left[ \langle \textbf{1}\textbf{2} \rangle 
	[\textbf{2} \textbf{1}] \langle \textbf{3} \lvert p_1 -p_2 \lvert \textbf{3}] + (\mathrm{cycl.})\right], \label{eq:WWZ}
	\\
	A_{WW\gamma}^{(\lambda_1\lambda_2\lambda_3)} 
	&= 
	\begin{tikzpicture}[baseline=(v)]
	\begin{feynman}[inline = (base.a), horizontal=p1 to v]
		\vertex (p1) [label=\({\scriptstyle W^+}\)];
		\vertex [below right = of p1] (v);
		\vertex [below left = of v,label=270:\({\scriptstyle W^-}\)] (p2);
		\vertex [right = of v, label=\({\scriptstyle \gamma}\)] (p3);
		\node [left = -0.25cm of v, blob, fill=gray] (vb);
		\begin{pgfonlayer}{bg}
		\diagram*{
		(p1) -- [photon,momentum=\({\scriptstyle p_1}\)] (v) -- [photon,rmomentum'=\({\scriptstyle p_2}\)] (p2),
		(v) -- [photon, rmomentum'=\({\scriptstyle p_3}\)] (p3),
		};
		\end{pgfonlayer}{bg}
	\end{feynman}
	\end{tikzpicture} = 
	\begin{cases}
	\displaystyle \frac{g_{WW\gamma}}{\sqrt{2}m_W^2 \langle 3 \xi_3\rangle}
	\left[\langle \mathbf{12}\rangle [\mathbf{21}] \langle {\xi_3 }\vert (p_1 - p_2) \vert 3]
	+ (\mathrm{cycl.})\right],
	&\mathrm{for}~~\lambda_3 = +,
	\vspace{1mm}
	\\
	\displaystyle -\frac{g_{WW\gamma}}{\sqrt{2}m_W^2 [ 3 \xi_3]}
	\left[\langle \mathbf{12}\rangle [\mathbf{21}] \langle {3 }\vert (p_1 - p_2) \vert \xi_3]
	+ (\mathrm{cycl.})\right],
	&\mathrm{for}~~\lambda_3 = -,
	\end{cases}
\end{align}
where we used the bold notation for the massive spinors, $\vert\xi_3\rangle, \vert \xi_3]$ represent the reference spinor of the photon, and ``(cycl.)" indicates the cyclic permutation of $1,2$ and $3$.
Note that the three-point amplitudes satisfy
\begin{align}
	&\left.A_{WWZ}^{(\lambda_1\lambda_2\lambda_3)}\right\vert_{\epsilon_k \to p_k}
	= -g_{WWZ}\frac{m_i^2 - m_j^2}{m_i m_j}\langle \mathbf{ij}\rangle [\mathbf{ji}],
	\quad
	\left.A_{WW\gamma}^{(\lambda_1\lambda_2\lambda_3)}\right\vert_{\epsilon_3 \to p_3}
	= 0,
	\label{eq:VVV_Ward}
\end{align}
where $i, j, k$ are understood to be cyclic, and the subscript means that 
we replace the polarization vector $\epsilon_k$ of the particle $k$ to its momentum $p_k$. For notational compactness, it is useful to define
\begin{align}
	V_{a\dot{a};b\dot{b};c\dot{c}}(p_1, p_2, p_3) &= 
	\frac{1}{4}
	\left[\epsilon_{ab}\epsilon_{\dot{a}\dot{b}} (p_1 - p_2)_{c\dot{c}}
	+ \epsilon_{bc}\epsilon_{\dot{b}\dot{c}} (p_2 - p_3)_{a\dot{a}}
	+ \epsilon_{ca}\epsilon_{\dot{c}\dot{a}} (p_3 - p_1)_{b\dot{b}}\right].
\end{align}
With this function, the three-point amplitudes are written as
\begin{align}
	A_{WWZ}^{(\lambda_1\lambda_2\lambda_3)}
	&= g_{WWZ} \epsilon_1^{\dot{a}{a}}\epsilon_2^{\dot{b}b}\epsilon_3^{\dot{c}c}
	V_{a\dot{a};b\dot{b};c\dot{c}}(p_1,p_2,p_3),
	\\
	A_{WW\gamma}^{(\lambda_1\lambda_2\lambda_3)}
	&= g_{WW\gamma} \epsilon_1^{\dot{a}{a}}\epsilon_2^{\dot{b}b}\epsilon_3^{\dot{c}c}
	V_{a\dot{a};b\dot{b};c\dot{c}}(p_1,p_2,p_3),
\end{align}
where we use a short-hand notation $\epsilon_i = \epsilon^{(\lambda_i)}(p_i)$.

%%%%%%%%%%%%%%%%%%%%%%%%%%%%%%%

%%%%%%%%%%%%%%%%%%%%%%%%%%%%%%%%%%%%%%%%%%

%%%%%%%%%%%%%%%%%%%%%%%%%%%%%%%%%%%%%%%
\subsection{$WWWW$ amplitude}
\label{subsec:EW_WWWW}
%%%%%%%%%%%%%%%%%%%%%%%%%%%%%%%%%%%%%%%

We begin with the $WWWW$ amplitude. 
We use the ALT shift to construct this amplitude and show that the contact term 
naturally emerges from recursion relations. 
Since this is one of our main results, we explain the calculation in some detail in the following. 

We label $W^+W^-W^+W^-$ respectively as 1,2,3,4, and take all the particles as incoming hereafter. In this process, we have the poles at
\begin{align}
	\hat{p}_{12}^2 = m_Z^2, \,\mathrm{or}~\,0,
	\quad
	\hat{p}_{14}^2 = m_Z^2, \,\mathrm{or}~\,0,
\end{align}
where we use the short-hand notation $\hat{p}_{ij} = \hat{p}_i + \hat{p}_j$.
We assume that the particle $1$ is transverse, $\lambda_1 = \pm$. 
The amplitude is constructible, and we take $B_\infty=0$.\footnote{One could take different points of view here. A common one is that one has to prove that the amplitude is constructible from other inputs, e.g., as shown in Sec.~\ref{subsec:ALTshift} with Ward identity. Another possibility is to \textit{assume} constructibility and study the behavior of the constructed from pole terms within a given shift and determine if it is compatible with the constructibility assumption. We defer the detailed discussion on these for future works.}
The amplitude can then be expressed solely in terms of the pole contributions (see Eq.~\eqref{eq:recursion_general})
\begin{align}
	A_{WWWW}^{(\lambda_1 \lambda_2 \lambda_3\lambda_4)}
	&= \sum_{I = Z, \gamma} \sum_{i=2,4}\frac{1}{p_{1i}^2-m_I^2}
	\frac{1}{{z^+_{1i} - z^-_{1i}}}
	\nonumber \\
	&\times\sum_{\lambda}
	\left[
	{z^+_{1i}} \hat{A}_{WWI}^{(\lambda_1 \lambda_i \lambda)}(z^-_{1i}) 
	\times \hat{A}_{WWI}^{(\lambda_3 \lambda_j\bar{\lambda})}(z^-_{1i})
	- {z^-_{1i}} \hat{A}_{WWI}^{(\lambda_1 \lambda_i \lambda)}(z^+_{1i}) 
	\times \hat{A}_{WWI}^{(\lambda_3 \lambda_j\bar{\lambda})}(z^+_{1i})
	\right],
	\label{eq:pole_WWWW}
\end{align}
where $j \neq i, 1, 3$ and we denote the solution of $\hat{p}_{ij}^2 = m_Z^2, 0$ as $z = z_{ij}^\pm$.\footnote{
	The solution $z= z_{ij}^\pm$ also depends on the mass of the intermediate particle $m_I^2$,
	but we do not put an explicit ``$I$"-index to $z_{ij}^\pm$ for notational simplicity.
} 

To simplify this, 
we use the completeness relation of the polarization vectors of the intermediate particle. The intermediate particles have polarization vectors whose little group indices are contracted,
which we indicate by  $\lambda$ and $\bar{\lambda}$,
and the contraction is expressed as
\begin{align}
	[\epsilon^{IJ}(\hat{p}_{1i})]_{a\dot{a}}[\epsilon_{IJ}(-\hat{p}_{1i})]_{b\dot{b}}
	&= -[\epsilon^{IJ}(\hat{p}_{1i})]_{a\dot{a}}[\epsilon_{IJ}(\hat{p}_{1i})]_{b\dot{b}}
	\nonumber \\
	&= \sum_{\lambda=\pm,L} [\epsilon^{(\lambda)}(\hat{p}_{1i})]_{a\dot{a}} [\epsilon^{(-\lambda)}(\hat{p}_{1i})]_{b\dot{b}}
	= -2\epsilon_{ab}\epsilon_{\dot{a}\dot{b}} + \frac{[\hat{p}_{1i}]_{a\dot{a}}[\hat{p}_{1i}]_{b\dot{b}}}{m_Z^2},
 \label{eq:internal_Z}
\end{align}
for the intermediate $Z$-boson,
where we used Eq.~\eqref{eq:spinor_negative_mom} in the first line and Eqs.~\eqref{eq:pol_massive_relation} and~\eqref{eq:pol_comp_massive_spinor} in the second line. Similarly, we have
\begin{align}
	[\epsilon^{IJ}(\hat{p}_{1i})]_{a\dot{a}}[\epsilon_{IJ}(-\hat{p}_{1i})]_{b\dot{b}}
	= -2\epsilon_{ab}\epsilon_{\dot{a}\dot{b}} 
	+ \frac{[\hat{p}_{1i}]_{a\dot{a}}[\hat{\bar{p}}_{1i}]_{b\dot{b}} 
	+ [\hat{\bar{p}}_{1i}]_{a\dot{a}}[\hat{p}_{1i}]_{b\dot{b}}}{\hat{p}_{1i}\cdot\hat{\bar{p}}_{1i}},
 \label{eq:internal_photon}
\end{align}
for the intermediate photon where $\bar{p}_i = \vert \xi_i\rangle [\xi_i\vert$. It then follows from Eq.~\eqref{eq:VVV_Ward} that we can write the product of the three-point amplitudes as
\begin{align}
	\sum_\lambda \hat{A}_{WWI}^{(\lambda_1 \lambda_i \lambda)} 
	\times \hat{A}_{WWI}^{(\lambda_3 \lambda_j\bar{\lambda})}
	= 
	-2g_{WWI}^2 \epsilon_1^{\dot{a}a}\epsilon_i^{\dot{b}b}
	V_{a\dot{a};b\dot{b};e\dot{e}}(\hat{p}_1,\hat{p}_i,-\hat{p}_{1i})
	{V_{c\dot{c};d\dot{d};}}^{\dot{e}e}(\hat{p}_3,\hat{p}_j,-\hat{p}_{3j})
	\epsilon_3^{\dot{c}c}\epsilon_j^{\dot{d}d},
\end{align}
where we do not have hats on the external polarization vectors since the ALT shift does not shift them. Since $V$ is linear in momentum, the product contains up to quadratic terms in $z_{1i}^\pm$ 
\begin{align}
	V_{a\dot{a};b\dot{b};e\dot{e}}(\hat{p}_1,\hat{p}_i,-\hat{p}_{1i})
	{V_{c\dot{c};d\dot{d};}}^{\dot{e}e}(\hat{p}_3,\hat{p}_j,-\hat{p}_{3j})
	&= V_{a\dot{a};b\dot{b};c\dot{c};d\dot{d}}^{(0)}
	+ V_{a\dot{a};b\dot{b};c\dot{c};d\dot{d}}^{(1)} z_{1i}^\pm
	+ V_{a\dot{a};b\dot{b};c\dot{c};d\dot{d}}^{(2)}\left(z_{1i}^\pm\right)^2.
\end{align}
Among these terms, $V^{(0)}$ is simply the product of the three-point amplitude without the momentum shift, while $V^{(1)}$ does not contribute to the final result 
since the contributions from $z_{1i}^\pm$ cancel with each other in Eq.~\eqref{eq:pole_WWWW}. In general, any linear dependence in $z$ will cancel after summing over all factorization channels.
The most interesting contribution is the quadratic term $V^{(2)}$. The overall $z_{1i}^\pm$ dependence of $V^{(2)}$ in Eq.~\eqref{eq:pole_WWWW} is given by
\begin{align}
	\frac{z_{1i}^+ (z_{1i}^-)^2 - z_{1i}^- (z_{1i}^+)^2}{z_{1i}^+ - z_{1i}^-}
	= -\frac{p_{1i}^2 - m_I^2}{2q_1 \cdot q_i}.
	\label{eq:z_square}
\end{align}
The numerator of this expression cancels the propagator $1/(p_{1i}^2 - m_I^2)$, and hence $V^{(2)}$ provides the contact term of the $WWWW$ scattering. We thus obtain
\begin{align}
	A_{WWWW}^{(\lambda_1 \lambda_2 \lambda_3\lambda_4)}
	=&
	-\sum_{I = Z, \gamma} \sum_{i=2,4}\frac{2g_{WWI}^2 }{p_{1i}^2-m_I^2}
	\epsilon_1^{\dot{a}a}\epsilon_i^{\dot{b}b}
	V_{a\dot{a};b\dot{b};e\dot{e}}({p}_1,{p}_i,-{p}_{1i})
	{V_{c\dot{c};d\dot{d};}}^{\dot{e}e}({p}_3,{p}_j,-{p}_{3j})
	\epsilon_3^{\dot{c}c}\epsilon_j^{\dot{d}d}
	\nonumber \\
	&+ \sum_{i=2,4}\frac{g^2}{q_1\cdot q_i}
	\epsilon_1^{\dot{a}a}\epsilon_i^{\dot{b}b}
	V_{a\dot{a};b\dot{b};e\dot{e}}({q}_1,{q}_i,-{q}_{1i})
	{V_{c\dot{c};d\dot{d};}}^{\dot{e}e}({q}_3,{q}_j,-{q}_{3j})
	\epsilon_3^{\dot{c}c}\epsilon_j^{\dot{d}d},
	\label{eq:4W_bfr_simplification}
\end{align}
where the first term comes from $V^{(0)}$ while the second term does from $V^{(2)}$, and we define
\begin{align}
	g^2 = g_{WWZ}^2 + g_{WW\gamma}^2.
	\label{eq:g2_square_sum}
\end{align}
To simplify the second line, we recall that we assume $\lambda_1 = \pm$, and hence the ALT shift takes the form
\begin{align}
	\epsilon_1 &= \frac{q_1}{c_1 m_W}.
\end{align}
We note that
\begin{align}
	q_1^{\dot{a}a}V_{a\dot{a};b\dot{b};e\dot{e}}(q_1, q_i, -q_{1i})
	&= (q_1\cdot q_i)\epsilon_{be}\epsilon_{\dot{b}\dot{e}}
	-\frac{1}{4}[q_i]_{b\dot{b}}[q_1]_{e\dot{e}}
	- \frac{1}{4}[q_1]_{b\dot{b}}(q_1+q_i)_{e\dot{e}}.
\end{align}
In this expression, the second term vanishes after contracting with the polarization vector $\epsilon_i$, while the third term vanishes after contracting with the other vertex function. This leaves us the first term, and we obtain
\begin{align}
	\sum_{i=2,4}\frac{\epsilon_1^{\dot{a}a}\epsilon_i^{\dot{b}b}}{q_1\cdot q_i}
	V_{a\dot{a};b\dot{b};e\dot{e}}({q}_1,{q}_i,-{q}_{1i})
	{V_{c\dot{c};d\dot{d};}}^{\dot{e}e}({q}_3,{q}_j,-{q}_{3j})
	\epsilon_3^{\dot{c}c}\epsilon_j^{\dot{d}d}
	&= \sum_{i=2,4}\frac{\epsilon_i^{\dot{b}b}}{c_1m_W}
	V_{c\dot{c};d\dot{d};b\dot{b}}(q_3,q_j,-q_{3j})\epsilon_3^{\dot{c}c}\epsilon_j^{\dot{d}d}.
\end{align}
With $\sum_i q_i = 0$ and $\epsilon_i \cdot q_i = 0$, we can simplify the last expression as
\begin{align}
	\sum_{i=2,4}\frac{\epsilon_i^{\dot{b}b}}{c_1m_W}
	V_{c\dot{c};d\dot{d};b\dot{b}}(q_3,q_j,-q_{3j})\epsilon_3^{\dot{c}c}\epsilon_j^{\dot{d}d}
	&=-\frac{1}{4} \epsilon_1^{\dot{a}a}\epsilon_3^{\dot{b}{b}}
	\left[2\epsilon_{ab}\epsilon_{\dot{a}\dot{b}}\epsilon_{cd}\epsilon_{\dot{c}\dot{d}}
	-\epsilon_{ac}\epsilon_{\dot{a}\dot{c}}\epsilon_{bd}\epsilon_{\dot{b}\dot{d}}
	-\epsilon_{ad}\epsilon_{\dot{a}\dot{d}}\epsilon_{bc}\epsilon_{\dot{b}\dot{c}}
	\right]
	\epsilon_2^{\dot{c}c}\epsilon_4^{\dot{d}d}.
\end{align}
We thus obtain
\begin{align}
	A_{WWWW}^{(\lambda_1 \lambda_2 \lambda_3\lambda_4)}
	=&
	-\sum_{I = Z, \gamma} \sum_{i=2,4}\frac{2g_{WWI}^2 }{p_{1i}^2-m_I^2}
	\epsilon_1^{\dot{a}a}\epsilon_i^{\dot{b}b}
	V_{a\dot{a};b\dot{b};e\dot{e}}({p}_1,{p}_i,-{p}_{1i})
	{V_{c\dot{c};d\dot{d};}}^{\dot{e}e}({p}_3,{p}_j,-{p}_{3j})
	\epsilon_3^{\dot{c}c}\epsilon_j^{\dot{d}d}
	\nonumber \\
	&-\frac{g^2}{4}\epsilon_1^{\dot{a}a}\epsilon_3^{\dot{b}{b}}
	\left[2\epsilon_{ab}\epsilon_{\dot{a}\dot{b}}\epsilon_{cd}\epsilon_{\dot{c}\dot{d}}
	-\epsilon_{ac}\epsilon_{\dot{a}\dot{c}}\epsilon_{bd}\epsilon_{\dot{b}\dot{d}}
	-\epsilon_{ad}\epsilon_{\dot{a}\dot{d}}\epsilon_{bc}\epsilon_{\dot{b}\dot{c}}
	\right]
	\epsilon_2^{\dot{c}c}\epsilon_4^{\dot{d}d}.
	\label{eq:EW_WWWW_final}
\end{align}
Its explicit form for all helicity combinations is given by
\begin{align}
	A_{WWWW}^{(\lambda_1 \lambda_2 \lambda_3\lambda_4)}
	=& -\sum_{I=Z,\gamma}\sum_{i=2,4}\frac{g^2_{WWI}}{ m_W^4(p_{1i}^2-m_I^2) } 
	\bigg[ 
	2{p}_1 \cdot ({p}_3 - {p}_j)
	\langle {\textbf{1}} {\textbf{i}}\rangle [{\textbf{i}}{\textbf{1}}] 
	\langle {\textbf{3}}{\textbf{j}}\rangle [{\textbf{j}}{\textbf{3}}]
 	\nonumber \\ 
	&
	- \langle {\textbf{1}} {\textbf{i}}\rangle [{\textbf{i}}{\textbf{1}}] 
	\big(
	\langle {\textbf{3}} \lvert ({p}_1 - {p}_i) \lvert {\textbf{3}}] \langle {\textbf{j}}\lvert {p}_3 \lvert {\textbf{j}}] 
	- \langle {\textbf{j}} \lvert 
	({p}_1 - {p}_i) \lvert {\textbf{j}}] \langle {\textbf{3}}\lvert {p}_j \lvert {\textbf{3}}]
	\big)
	+ (({1},{i})\leftrightarrow ({3},{j}))
	\nonumber \\
 	&
	+2 \langle {\textbf{1}} {\textbf{3}}\rangle [{\textbf{3}} {\textbf{1}}] 
	\langle {\textbf{i}}\lvert {p}_1 \lvert {\textbf{i}}] \langle {\textbf{j}} \lvert {p}_3 \lvert {\textbf{j}}] 
	- 2 \langle {\textbf{1}} {\textbf{j}}\rangle [{\textbf{j}} {\textbf{1}}] 
	\langle {\textbf{i}}\lvert {p}_1 \lvert {\textbf{i}}] \langle {\textbf{3}} \lvert {p}_j \lvert {\textbf{3}}] 
	- ({1} \leftrightarrow {i}) 
	\bigg]
	\nonumber \\
      	&- \frac{g^2}{m_W^4 } \bigg[
	2 \langle \textbf{24}\rangle [\textbf{42}]  \langle \textbf{13} \rangle [\textbf{31}] 
	-  \langle \textbf{23}\rangle [\textbf{32}]  \langle \textbf{14} \rangle [\textbf{41}] 
	-  \langle \textbf{12}\rangle [\textbf{21}]  \langle \textbf{34} \rangle [\textbf{43}]
	\bigg],
\end{align}
where we use the bold notation for the polarization spinors. This is our result of the $WWWW$ scattering based on the on-shell recursion relation. Although the amplitude was constructed assuming $\lambda_1 = \pm$, 
the little-group covariance and spin-raising/lowering operators allow us to write the amplitude for all polarizations. 
We check and find that this agrees with the Feynman diagrammatic result.

We summarize several properties of our computation:
\begin{itemize}

\item The final result does not depend on $q_i$. This is expected since the original amplitude does not depend on how we shift the momentum. Therefore, the $q_i$-independence serves as a consistency check
of our argument that $B_\infty = 0$. Indeed, our simplification of the $z^2$ term is analogous to the Ward identity in
the massless theory.

\item We checked explicitly that the $q_i$-dependence does not cancel for all longitudinal scattering when directly summing the pole contributions with ALT shift. \footnote{While the additional degrees of freedom in $c_i$ could be further utilized to remove the explicit $q_i$ dependence for this amplitude, this general $q_i$ dependence, which is outside the allowed components (polarization vectors and momentum vectors) of the physical amplitudes for all-longitudinal scattering, indicates the constructibility issue with the vanilla ALT shift under study for this particular amplitude.} This is consistent with our discussion in Sec.~\ref{subsec:ALTshift}, that we cannot guarantee the constructibility of this amplitude with our shift through dimensional analysis, even with additional external input. Instead, one can obtain the correct longitudinal amplitude through spin raising and lowering operators.

\item We do not need an explicit form of $c_i$. The square root of the momentum products in $z_{1i}^\pm$ also cancels since the poles always appear in pairs.

\item The contact term arises from the $z^2$ term in the product of the three-point amplitudes, which in turn arises from the parts dependent quadratically on the external momenta. Therefore, we can reconstruct this term \emph{only if} we introduce the explicit momentum shift such as the ALT shift.\footnote{Note that Ref.~\cite{Christensen:2024xzs} also got the contact term by dropping terms depending on $\hat s$, which is a process allowed by consistent shift but not necessarily uniquely yield the correct amplitude. It would be interesting to see how far along this method can go for higher point amplitudes or general theories, while the consistent ALT shift introduced here is guaranteed to apply to general theories.}

\end{itemize}
In particular, we believe that we, for the first time, have derived the contact term of the $WWWW$ scattering explicitly in the on-shell recursion relation.

%%%%%%%%%%%%%%%%%%%

%%%%%%%%%%
\subsubsection*{High-energy limit}
%%%%%%%%%%

As we have noted, even though the ALT shift by itself does not allow 
the construction of the all-longitudinal $WWWW$ scattering,
we can exploit the little-group covariance to construct it from the amplitude with (at least) one transverse mode.
We can then investigate the high-energy limit of the all-longitudinal $WWWW$ amplitude.

We take the high-energy limit with a fixed scattering angle and let $s, t \to \infty$, where we define the Mandelstam
variables $(s,t,u) = (p_{12}^2, p_{13}^2, p_{14}^2)$.
In this limit, the amplitude with all longitudinal modes behaves as
\begin{align}
	\left.A_{WWWW}^{(LLLL)}\right\vert_{Z,\gamma}
	=& -\left(g^2 - g_{WWZ}^2 - g_{WW\gamma}^2\right)\frac{s^2 + 4su + u^2}{4m_W^4}
	\nonumber \\
	&-\frac{s+u}{4 m_W^4}
	\left[
	4(2g_{WWZ}^2 + 2g_{WW\gamma}^2-g^2)m_W^2 - 3 g_{WWZ}^2 m_Z^2
	+ \frac{8 m_W^2 u}{s} (g^2 - g_{WWZ}^2 - g_{WW\gamma}^2)
	\right]
	\nonumber \\
	&+ \mathcal{O}(E^0),
\end{align}
where we make it explicit that Eq.~\eqref{eq:g2_square_sum} guarantees the cancellation of the $\mathcal{O}(E^4)$ term,
and the subscript indicates that this amplitude is from the intermediate $Z$ and $\gamma$ only (and no Higgs).
After using Eq.~\eqref{eq:g2_square_sum}, which we derived using the ALT shift, it is simplified as
\begin{align}
	\left.A_{WWWW}^{(LLLL)}\right\vert_{Z,\gamma}
	&= -\frac{s+u}{4 m_W^4}
	(4g^2m_W^2 - 3 g_{WWZ}^2 m_Z^2) + \mathcal{O}(E^0).
\end{align}
Therefore the $\mathcal{O}(E^4)$ term cancels automatically in our recursive construction of the amplitudes, 
while the $\mathcal{O}(E^2)$ cancellation requires additional input, such as an introduction of the Higgs,
which is the main topic of Sec.~\ref{sec:UV}.

%%%%%%%%%%%%%%%%%%%%%%%%%%%%%%%%%%%%%%%
\subsection{$WWZZ$ amplitude}
%%%%%%%%%%%%%%%%%%%%%%%%%%%%%%%%%%%%%%%

We next consider the $WWZZ$ scattering. The computation is analogous to the $WWWW$ scattering,
and therefore, we omit details.
Labelling $W^+,W^-,Z, Z$ as $1,2,3,4$, we have the poles at
\begin{align}
	\hat{p}_{13}^2 = m_W^2,
	\quad
	\hat{p}_{14}^2 = m_W^2.
\end{align}
We take the particle $1$ as transverse, $\lambda_1 = \pm$, which guarantees $B_\infty = 0$ with ALT shift, and write
\begin{align}
	A_{WWZZ}^{(\lambda_1 \lambda_2 \lambda_3\lambda_4)}
	&= \sum_{i=3,4}\frac{1}{p_{1i}^2-m_W^2}
	\frac{1}{{z^+_{1i} - z^-_{1i}}}
	\nonumber \\
	&\times\sum_{\lambda}
	\left[
	{z^+_{1i}} \hat{A}_{WWZ}^{(\lambda_1\lambda \lambda_i )}(z^-_{1i}) 
	\times \hat{A}_{WWZ}^{(\bar{\lambda}\lambda_2 \lambda_j)}(z^-_{1i})
	- {z^-_{1i}} \hat{A}_{WWZ}^{(\lambda_1\lambda\lambda_i)}(z^+_{1i}) 
	\times \hat{A}_{WWZ}^{(\bar{\lambda}\lambda_2 \lambda_j)}(z^+_{1i})
	\right].
	\label{eq:pole_WWZZ}
\end{align}
 We use the completeness relation of the intermediate polarization vector,
\begin{align}
	[\epsilon^{IJ}(\hat{p}_{1i})]_{a\dot{a}}[\epsilon_{IJ}(-\hat{p}_{1i})]_{b\dot{b}}
	&= -2\epsilon_{ab}\epsilon_{\dot{a}\dot{b}} + \frac{[\hat{p}_{1i}]_{a\dot{a}}[\hat{p}_{1i}]_{b\dot{b}}}{m_W^2}.
 \label{eq:vec_completeness}
\end{align}
In the present case, the second term gives 
a finite contribution as opposed to the $WWWW$ scattering.
With Eq.~\eqref{eq:VVV_Ward}, this contribution does not change under the momentum shift,
and hence we can drop the hats from the momentum for this term.
Therefore, the $z^2$-term arises solely from the first term,
and with a similar computation as the $WWWW$ scattering, we obtain
\begin{align}
	A_{WWZZ}^{(\lambda_1 \lambda_2 \lambda_3\lambda_4)}
	=&
	-\sum_{i=3,4}\frac{2g_{WWZ}^2 }{p_{1i}^2-m_W^2}
	\epsilon_1^{\dot{a}a}\epsilon_i^{\dot{b}b}
	V_{a\dot{a};e\dot{e};b\dot{b}}({p}_1,-{p}_{1i},p_i)
	\left[\epsilon^{ef}\epsilon^{\dot{e}\dot{f}}
	-\frac{[p_{1i}]^{\dot{e}e}[p_{1i}]^{\dot{f}f}}{2m_W^2}
	\right]
	V_{f\dot{f};c\dot{c};d\dot{d}}(-{p}_{2j},{p}_2,{p}_j)
	\epsilon_2^{\dot{c}c}\epsilon_j^{\dot{d}d}
	\nonumber \\
	&+ \sum_{i=3,4}\frac{g_{WWZ}^2}{q_1\cdot q_i}
	\epsilon_1^{\dot{a}a}\epsilon_i^{\dot{b}b}
	V_{a\dot{a};e\dot{e};b\dot{b}}({q}_1,-{q}_{1i},{q}_i)
	{V^{\dot{e}e;}}_{c\dot{c};d\dot{d}}(-q_{2j}, {q}_2,{q}_j)
	\epsilon_2^{\dot{c}c}\epsilon_j^{\dot{d}d},
	\label{eq:WWZZ_bfr_simplification}
\end{align}
where $j\neq 1,2,i$. Again, the term in the second line arising from $z^2$ provides the contact term.
As before, we can simplify it with the relation $\epsilon_1 = q_1/c_1 m_W$, and we obtain
\begin{align}
	A_{WWZZ}^{(\lambda_1 \lambda_2 \lambda_3\lambda_4)}
	=&
	-\sum_{i=3,4}\frac{2g_{WWZ}^2 }{p_{1i}^2-m_W^2}
	\epsilon_1^{\dot{a}a}\epsilon_i^{\dot{b}b}
	V_{a\dot{a};e\dot{e};b\dot{b}}({p}_1,-{p}_{1i},p_i)
	\left[\epsilon^{ef}\epsilon^{\dot{e}\dot{f}}
	-\frac{[p_{1i}]^{\dot{e}e}[p_{1i}]^{\dot{f}f}}{2m_W^2}
	\right]
	V_{f\dot{f};c\dot{c};d\dot{d}}(-{p}_{2j},{p}_2,{p}_j)
	\epsilon_2^{\dot{c}c}\epsilon_j^{\dot{d}d}
	\nonumber \\
	&+\frac{g_{WWZ}^2}{4}\epsilon_1^{\dot{a}a}\epsilon_2^{\dot{b}{b}}
	\left[2\epsilon_{ab}\epsilon_{\dot{a}\dot{b}}\epsilon_{cd}\epsilon_{\dot{c}\dot{d}}
	-\epsilon_{ac}\epsilon_{\dot{a}\dot{c}}\epsilon_{bd}\epsilon_{\dot{b}\dot{d}}
	-\epsilon_{ad}\epsilon_{\dot{a}\dot{d}}\epsilon_{bc}\epsilon_{\dot{b}\dot{c}}
	\right]
	\epsilon_3^{\dot{c}c}\epsilon_4^{\dot{d}d}.
	\label{eq:EW_WWZZ_final}
\end{align}
Its explicit form is given by
\begin{align}
	A_{WWZZ}^{(\lambda_1 \lambda_2 \lambda_3\lambda_4)}
	=&- \sum_{i=3,4}\frac{g^2_{WWZ}}{ m_W^2 m_Z^2(p_{1i}^2-m_W^2)}  
	\bigg[\left(2({p}_1 \cdot {p}_2 -{p}_1 \cdot {p}_j)-\frac{(m_W^2 - m_Z^2)^2}{m_W^2}\right) 
	\langle {\textbf{1i}}\rangle [{\textbf{i1}}] \langle {\textbf{2j}}\rangle [{\textbf{j2}}]
	\nonumber \\
 	&- \langle {\textbf{1i}}\rangle [{\textbf{i1}}] \langle {\textbf{2}} \lvert ({p}_1 - {p}_i) \lvert {\textbf{2}}] 
	\langle {\textbf{j}}\lvert {p}_2 \lvert {\textbf{j}}] +  \langle {\textbf{1i}}\rangle [{\textbf{i1}}] \langle {\textbf{j}} \lvert ({p}_1 - {p}_i) \lvert {\textbf{j}}] \langle {\textbf{2}}\lvert {p}_j \lvert {\textbf{2}}] + ({1},{i})\leftrightarrow ({2},{j}) \nonumber \\
 &  +2 \langle {\textbf{12}}\rangle [{\textbf{21}}] \langle {\textbf{i}}\lvert {p}_1 \lvert {\textbf{i}}] \langle {\textbf{j}} \lvert {p}_2 \lvert {\textbf{j}}] - 2 \langle {\textbf{1j}}\rangle [{\textbf{j1}}] \langle {\textbf{i}}\lvert {p}_1 \lvert {\textbf{i}}] \langle {\textbf{2}} \lvert {p}_j \lvert {\textbf{2}}] - ({1} \leftrightarrow {i}) \bigg],
 	\nonumber \\
	&+ \frac{g_{WWZ}^2}{m_W^2 m_Z^2} \bigg[
	2\langle \textbf{12} \rangle [\textbf{21}] \langle \textbf{34}\rangle [\textbf{43}]  
	- \langle \textbf{14} \rangle [\textbf{41}] \langle \textbf{23}\rangle [\textbf{32}]  
	-\langle \textbf{13}\rangle [\textbf{31}]  \langle \textbf{24} \rangle [\textbf{42}]   
	\bigg].
\end{align}
This is our result of the $WWZZ$ scattering.

%%%%%%%%%%
\subsubsection*{High-energy limit}
%%%%%%%%%%

The high-energy limit of the $WWZZ$ amplitude is given by
\begin{align}
     \left.A_{WWZZ}^{(LLLL)}\right\vert_{W}= \frac{g_{WWZ}^2m_Z^2(t+u)}{4 m_W^4} + \mathcal{O}(E^0).
\end{align}
In particular, the $\mathcal{O}(E^4)$ term again automatically vanishes in our on-shell computation,
while the $\mathcal{O}(E^2)$ term needs additional input.

%%%%%%%%%%%%%%%%%%%%%%%%%%%%%%%%%%%%%%%
\subsection{$WWtt$ amplitude}
%%%%%%%%%%%%%%%%%%%%%%%%%%%%%%%%%%%%%%%
In this subsection, we illustrate the on-shell construction of four-point massive gauge boson and fermion amplitude with dimensionless couplings. We will explicitly work out $WWtt$ amplitude as an example, but the discussion is general and it can be extended to other processes. For this, we introduce one generation of quarks, namely $t$ and $b$ quarks with masses $m_t$ and $m_b$, and assume their electric charges are $+2/3$ and $-1/3$ respectively. With these additional particles, we can write down the following minimal three-point interaction

\begin{align}
	A_{tbW}^{(\lambda_1\lambda_2\lambda_3)}
	&= 
	\begin{tikzpicture}[baseline=(v)]
	\begin{feynman}[inline = (base.a), horizontal=p1 to v]
		\vertex (p1) [label=\({\scriptstyle t}\)];
		\vertex [below right = of p1] (v);
		\vertex [below left = of v,label=270:\({\scriptstyle \bar{b}}\)] (p2);
		\vertex [right = of v, label=\({\scriptstyle W^-}\)] (p3);
		\node [left = -0.25cm of v, blob, fill=gray] (vb);
		\begin{pgfonlayer}{bg}
		\diagram*{
		(p1) -- [fermion,momentum=\({\scriptstyle p_1}\)] (v) -- [fermion,rmomentum'=\({\scriptstyle p_2}\)] (p2),
		(v) -- [photon, rmomentum'=\({\scriptstyle p_3}\)] (p3),
		};
		\end{pgfonlayer}{bg}
	\end{feynman}
	\end{tikzpicture}
	= \frac{\sqrt{2}g_{tbW}}{m_W}  \langle \textbf{1}\textbf{3} \rangle 
	[\textbf{3} \textbf{2}],
	\\
	A_{ttZ}^{(\lambda_1\lambda_2\lambda_3)}
	&= 
	\begin{tikzpicture}[baseline=(v)]
	\begin{feynman}[inline = (base.a), horizontal=p1 to v]
		\vertex (p1) [label=\({\scriptstyle {t}}\)];
		\vertex [below right = of p1] (v);
		\vertex [below left = of v,label=270:\({\scriptstyle \bar{t}}\)] (p2);
		\vertex [right = of v, label=\({\scriptstyle Z}\)] (p3);
		\node [left = -0.25cm of v, blob, fill=gray] (vb);
		\begin{pgfonlayer}{bg}
		\diagram*{
		(p1) -- [fermion,momentum=\({\scriptstyle p_1}\)] (v) -- [fermion,rmomentum'=\({\scriptstyle p_2}\)] (p2),
		(v) -- [photon, rmomentum'=\({\scriptstyle p_3}\)] (p3),
		};
		\end{pgfonlayer}{bg}
	\end{feynman}
	\end{tikzpicture}
	= \frac{\sqrt{2}}{m_W} \left[ g_{ttZ}^L \langle \textbf{1}\textbf{3} \rangle [\textbf{3}\textbf{2}]+ g_{ttZ}^R [ \textbf{1}\textbf{3} ]
	\langle \textbf{3} \textbf{2}\rangle \right],
	\\
 A_{tt\gamma}^{(\lambda_1\lambda_2\lambda_3)} 
	&= 
	\begin{tikzpicture}[baseline=(v)]
	\begin{feynman}[inline = (base.a), horizontal=p1 to v]
		\vertex (p1) [label=\({\scriptstyle t}\)];
		\vertex [below right = of p1] (v);
		\vertex [below left = of v,label=270:\({\scriptstyle \bar{t}}\)] (p2);
		\vertex [right = of v, label=\({\scriptstyle \gamma}\)] (p3);
		\node [left = -0.25cm of v, blob, fill=gray] (vb);
		\begin{pgfonlayer}{bg}
		\diagram*{
		(p1) -- [fermion,momentum=\({\scriptstyle p_1}\)] (v) -- [fermion,rmomentum'=\({\scriptstyle p_2}\)] (p2),
		(v) -- [photon, rmomentum'=\({\scriptstyle p_3}\)] (p3),
		};
		\end{pgfonlayer}{bg}
	\end{feynman}
	\end{tikzpicture}
	= 
	\begin{cases}
	\displaystyle \frac{\sqrt{2}g_{tt\gamma}}{\langle 3 \xi_3\rangle} \bigg[
	\langle \mathbf{2} \xi_3\rangle [3 \mathbf{1}] + \langle \mathbf{1} \xi_3\rangle [3 \mathbf{2}]\bigg] ,
	&\mathrm{for}~~\lambda_3 = +,
	\vspace{1mm}
	\\
	\displaystyle - \frac{\sqrt{2}g_{tt\gamma}}{[ 3 \xi_3 ]}\bigg[  [ \mathbf{2} \xi_3] \langle 3 \mathbf{1} \rangle+[ \mathbf{1} \xi_3] \langle 3 \mathbf{2} \rangle \bigg],
	&\mathrm{for}~~\lambda_3 = -.
	\end{cases}
\end{align}
Here we have added parity violating interactions to align with the Standard Model, but on-shell constructibiliy is independent of parity violation of these three-point amplitudes. 
Now let us denote $W^+,W^-,{t}, \bar{t}$ as $1,2,3,4$, then the poles of the factorization channels are
\begin{align}
    \hat{p}_{12}^2 = 0,\,~\mathrm{or}~\,m_Z^2, \quad    \hat{p}_{14}^2 = m_b^2.
\end{align}
and the amplitude is given by
\begin{align}
	A_{WWtt}^{(\lambda_1 \lambda_2 \lambda_3 \lambda_4)}
	&= \sum_{I=Z,\gamma} \frac{1}{p_{12}^2-m_I^2}
	\frac{1}{{z^+_{12} - z^-_{12}}} \nonumber \\
 &\times
	\sum_{\lambda}
	\left[
	{z^+_{12}} \hat{A}_{WWI}^{(\lambda_1\lambda_2 \lambda )}(z^-_{12}) 
	\times \hat{A}_{ttI}^{(\lambda_3\lambda_4 \bar{\lambda})}(z^-_{12})
	- {z^-_{12}} \hat{A}_{WWI}^{(\lambda_1\lambda_2 \lambda)}(z^+_{12}) 
	\times \hat{A}_{ttI}^{(\lambda_3\lambda_4 \bar{\lambda})}(z^+_{12})
	\right]
	\nonumber \\
	&+\frac{1}{p_{14}^2-m_b^2}
	\frac{1}{{z^+_{14} - z^-_{14}}} \nonumber\\
 &\times
	\sum_{\lambda}
	\left[
	{z^+_{14}} {\hat{A}}^{(\lambda_4 \lambda \lambda_1)}_{tbW}(z^-_{14}) 
	\times \hat{A}_{tbW}^{(\lambda_3 \bar{\lambda} \lambda_2)}(z^-_{14})
	- {z^-_{14}} \hat{A}_{tbW}^{(\lambda_4\lambda\lambda_1)}(z^+_{14}) 
	\times \hat{A}_{tbW}^{(\lambda_3 \bar{\lambda} \lambda_2 )}(z^+_{14})
	\right].
	\label{eq:pole_WWtt}
\end{align}
In contrast to the four gauge boson scatterings, here we do not assume that $\lambda_1 = \pm$
since the external fermions improve the large-$z$ behavior; see Eq.~\eqref{eq:large_z_Ward}.
This amplitude has internal massive fermion and vector, and we use the completeness relation for fermion Eq.~\eqref{eq:fermion_complt} and vectors in Eq.~(\ref{eq:internal_Z}). Then, we can rewrite
\begin{align}
 \sum_{\lambda}   \hat{A}_{WWI}^{(\lambda_1\lambda_2 \lambda )}
	\times \hat{A}_{ttI}^{(\lambda_3\lambda_4 \bar{\lambda})} &= -2g_{WWI} \epsilon_1^{\dot{a}a}\epsilon_2^{\dot{b}b}{V_{a\dot{a};b\dot{b}}}^{c\dot{c}}(\hat{p}_1,\hat{p}_{2},-\hat{p}_{12}) (g_{ttI}^L \lvert\textbf{3}\rangle_c[\textbf{4}\lvert_{\dot{c}}+ g_{ttI}^R\lvert\textbf{4}\rangle_c[\textbf{3}\lvert_{\dot{c}}),  \\
 \sum_{\lambda} \hat{A}_{tbW}^{(\lambda_4 \lambda \lambda_1 )}
	\times \hat{A}_{tbW}^{(\lambda_3 \bar{\lambda} \lambda_2)}
	&=- \frac{2g_{tbW}^2}{m_W^2} [ {\textbf{4}} {\textbf{1}}] \langle {\textbf{2}} {\textbf{3}}\rangle \langle \textbf{1}\lvert \hat{p}_{14}  \lvert {\textbf{2}}],
 \end{align}
where we defined $g_{tt\gamma}^L = g_{tt\gamma}^R = g_{tt\gamma}$. Now, we notice that all terms in Eq.~\eqref{eq:pole_WWtt} have one momentum insertion at maximum in the numerator which originates from  ${V_{a\dot{a};b\dot{b}}}^{c\dot{c}}(\hat{p}_1,\hat{p}_{2},-\hat{p}_{12})$ or internal fermion. 
They contain only up to linear terms in $z$, and the linear terms cancel after summing over the poles of each factorization channel. Therefore, in contrast to the $WWWW$ and $WWZZ$ amplitudes, the $WWtt$ amplitude does not have a four-point contact term. Then, we can write the full amplitude in spinor representation as
\begin{align}
    A_{WWtt}^{(\lambda_1 \lambda_2 \lambda_3 \lambda_4)} = &- \sum_{I=Z,\gamma} \frac{1}{m_W^2(p_{12}^2-m_I^2)}  \bigg( g_{WWI}g_{ttI}^L ( \langle {\textbf{1}} {\textbf{2}} \rangle [{\textbf{2}} {\textbf{1}}]  \langle {\textbf{3}} \lvert {p}_1 -{p}_2 \lvert {\textbf{4}}]-2 \langle {\textbf{3}}{\textbf{1}}\rangle [{\textbf{1}}{\textbf{4}}] \langle {\textbf{2}} \lvert {p}_1 \lvert {\textbf{2}}] \nonumber  \\
    &+2\langle {\textbf{3}}{\textbf{2}}\rangle [{\textbf{2}}{\textbf{4}}] \langle {\textbf{1}} \lvert {p}_2 \lvert {\textbf{1}}] ) 
     + g_{WWI}g_{ttI}^R ( \langle {\textbf{1}} {\textbf{2}} \rangle [{\textbf{2}} {\textbf{1}}]  \langle {\textbf{4}} \lvert {p}_1 -{p}_2 \lvert {\textbf{3}}]-2 \langle {\textbf{4}}{\textbf{1}}\rangle [{\textbf{1}}{\textbf{3}}] \langle {\textbf{2}} \lvert {p}_1 \lvert {\textbf{2}}] \nonumber \\
     &+2\langle {\textbf{4}}{\textbf{2}}\rangle [{\textbf{2}}{\textbf{3}}] \langle {\textbf{1}} \lvert {p}_2 \lvert {\textbf{1}}] ) \bigg)  -\frac{2 g_{tbW}^2}{m_W^2 (p_{14}^2 - m_b^2)}  [ {\textbf{4}} {\textbf{1}}] \langle {\textbf{2}} {\textbf{3}} \rangle \langle \textbf{1}\lvert {p}_{14}  \lvert {\textbf{2}}] .
     \label{eq:WWtt_spin_rep}
\end{align}

%%%%%%%%%%%%%%%%%%%%%%%%%%%%%%%%%%%%%%%
\section{UV completion of electroweak theory}
\label{sec:UV}
%%%%%%%%%%%%%%%%%%%%%%%%%%%%%%%%%%%%%%%

In Sec.~\ref{sec:EW}, we have seen that we can construct the EW gauge boson amplitudes,
including the contact terms, with the ALT shift even in the absence of the Higgs boson.
There we see that our calculation results in the automatic cancellation of the $\mathcal{O}(E^4)$ term
in the all-longitudinal EW gauge boson amplitude at high energy,
which leaves us only the $\mathcal{O}(E^2)$ terms.

Canceling the $\mathcal{O}(E^2)$ terms requires additional input.
Here, we have different options, e.g., we can introduce new particles
such as scalar and vector particles with different interactions. The main topic of this section is to study these UV completions in the on-shell method, again with the ALT shift.
We mainly focus on UV completion by scalar particles, corresponding to Higgs multiplets,
and only briefly mention UV completion by vector particles at the end. We can see that studying the unitarity behavior will further restrict relations between the coefficients of different three-point amplitudes.

%%%%%%%%%%%%%%%%%%%%%%%%%%%%%%%%%%%%%%%
\subsection{A neutral scalar: Higgs doublet}
\label{subsec:UV_doublet}
%%%%%%%%%%%%%%%%%%%%%%%%%%%%%%%%%%%%%%%

We begin with adding one real scalar particle, $h$, with its mass $m_h$, 
which we call the Higgs. 
Indeed, after requiring the cancellations of the $\mathcal{O}(E^2)$ terms,
it will be clear that this naturally leads to UV completion by a Higgs doublet.

The minimal three-point amplitudes of a real scalar particle and vector particles are given by
\begin{align}
	A_{WWh}^{(\lambda_1\lambda_2)}
	&= \begin{tikzpicture}[baseline=(v)]
	\begin{feynman}[inline = (base.a), horizontal=p1 to v]
		\vertex (p1) [label=\({\scriptstyle W^\pm}\)];
		\vertex [below right = of p1] (v);
		\vertex [below left = of v,label=270:\({\scriptstyle W^\mp}\)] (p2);
		\vertex [right = of v, label=\({\scriptstyle h}\)] (p3);
		\node [left = -0.25cm of v, blob, fill=gray] (vb);
		\begin{pgfonlayer}{bg}
		\diagram*{
		(p1) -- [photon,momentum=\({\scriptstyle p_1}\)] (v) -- [photon,rmomentum'=\({\scriptstyle p_2}\)] (p2),
		(v) -- [scalar, rmomentum'=\({\scriptstyle p_3}\)] (p3),
		};
		\end{pgfonlayer}{bg}
	\end{feynman}
	\end{tikzpicture}
	= \frac{g_{WWh}}{ m_W^2} \langle \textbf{1}\textbf{2} \rangle [\textbf{2}\textbf{1}],
	\quad
	A_{ZZh}^{(\lambda_1\lambda_2)}
	= 
	\begin{tikzpicture}[baseline=(v)]
	\begin{feynman}[inline = (base.a), horizontal=p1 to v]
		\vertex (p1) [label=\({\scriptstyle Z}\)];
		\vertex [below right = of p1] (v);
		\vertex [below left = of v,label=270:\({\scriptstyle Z}\)] (p2);
		\vertex [right = of v, label=\({\scriptstyle h}\)] (p3);
		\node [left = -0.25cm of v, blob, fill=gray] (vb);
		\begin{pgfonlayer}{bg}
		\diagram*{
		(p1) -- [photon,momentum=\({\scriptstyle p_1}\)] (v) -- [photon,rmomentum'=\({\scriptstyle p_2}\)] (p2),
		(v) -- [scalar, rmomentum'=\({\scriptstyle p_3}\)] (p3),
		};
		\end{pgfonlayer}{bg}
	\end{feynman}
	\end{tikzpicture}
	= \frac{g_{ZZh}}{ m_Z^2} \langle \textbf{1}\textbf{2} \rangle [\textbf{2}\textbf{1}].
\end{align}
In the on-shell method, 
$g_{WWh}$ and $g_{ZZh}$ are arbitrary at this point,
and we will fix it by requiring the cancellation of the $\mathcal{O}(E^2)$ terms.
On top of these couplings, we have three-point amplitudes of the Higgs particles,
but they do not lead to $\mathcal{O}(E^2)$ terms.
Therefore these contributions are out of our interest, and we ignore them in the rest part.
In the following, we construct four-point amplitudes from these three-point amplitudes using the on-shell method.

%%%%%%%%%%
\subsubsection*{$WWWW$ amplitude}
%%%%%%%%%%

We start from the $WWWW$ amplitude. The Higgs contribution is given by
\begin{align}
	\left.A_{WWWW}^{(\lambda_1 \lambda_2 \lambda_3\lambda_4)}\right\vert_h
	&= \sum_{i=2,4}\frac{1}{p_{1i}^2-m_h^2}
	\frac{1}{{z^+_{1i} - z^-_{1i}}}
	\nonumber \\
	&\times
	\left[
	{z^+_{1i}} \hat{A}_{WWh}^{(\lambda_1 \lambda_i)}(z^-_{1i}) 
	\times \hat{A}_{WWh}^{(\lambda_3 \lambda_j)}(z^-_{1i})
	- {z^-_{1i}} \hat{A}_{WWh}^{(\lambda_1 \lambda_i)}(z^+_{1i}) 
	\times \hat{A}_{WWh}^{(\lambda_3 \lambda_j}(z^+_{1i})
	\right],
	\label{eq:pole_WWWW_higgs}
\end{align}
where $z_{1i}^{\pm}$ are the roots for the pole condition $\hat{p}_{1i}^2 = m_h^2$. The three-point amplitudes do not depend on $z$ since the polarization vectors are not shifted,
and therefore it is simplified as the product of the three-point amplitudes without the momentum shift as
\begin{align}
	\left.A_{WWWW}^{(\lambda_1 \lambda_2 \lambda_3\lambda_4)}\right\vert_h
	&= \frac{g_{WWh}^2}{m_W^4}
	\left[
	\frac{\langle \mathbf{12}\rangle [\mathbf{21}]\langle \mathbf{34}\rangle [\mathbf{43}]}
	{p_{12}^2 - m_{h}^2}
	+ \frac{\langle \mathbf{14}\rangle [\mathbf{41}]\langle \mathbf{32}\rangle [\mathbf{23}]}
	{p_{14}^2 - m_{h}^2}
	\right].
	\label{eq:WWWW_higgs_doublet}
\end{align}
Incorporating this contribution, the high energy limit of the all-longitudinal amplitude is given by
\begin{align}
	A_{WWWW}^{(LLLL)}
	&= -\frac{s+u}{4 m_W^4}
	\left[4g^2m_W^2 - 3 g_{WWZ}^2 m_Z^2-g_{WWh}^2
	\right] + \mathcal{O}(E^0).
\end{align}
Therefore good high-energy behavior of this channel requires
\begin{align}
	g^2_{WWh} = 4 g^2 m_W^2 - 3 g_{WWZ}^2 m_Z^2.
	\label{eq:constraint1}
\end{align}
%%

%%%%%%%%%%
\subsubsection*{$WWZZ$ amplitude}
%%%%%%%%%%
The computation of the $WWZZ$ scattering is analogous to the $WWWW$ scattering, and we obtain
\begin{align}
	\left.A_{WWZZ}^{(\lambda_1 \lambda_2 \lambda_3\lambda_4)}\right\vert_h
	&= \frac{g_{WWh} g_{ZZh}}{m_W^2 m_Z^2}
	\frac{\langle \mathbf{12}\rangle [\mathbf{21}]\langle \mathbf{34}\rangle [\mathbf{43}]}
	{p_{12}^2 - m_{h}^2}.
	\label{eq:WWZZ_higgs_doublet}
\end{align}
Including this contribution, the high-energy limit of the all-longitudinal amplitude is given by
\begin{align}
	A_{WWZZ}^{(LLLL)}
	&= \frac{t+u}{4m_W^2 m_Z^2}\left[\frac{g_{WWZ}^2 m_Z^4}{m_W^2} 
	- g_{WWh}  g_{ZZh}
	\right]
	+ \mathcal{O}(E^0).
\end{align}
Therefore UV completion of this channel requires
\begin{align}
	g_{WWh} g_{ZZh} &= \frac{g_{WWZ}^2 m_Z^4}{m_W^2}.
	\label{eq:constraint2}
\end{align}
We also have the $ZZZZ$ scattering, in the presence of the Higgs, 
but it grows only as $\mathcal{O}(E^0)$ due to $s + t + u = 4m_Z^2$,
and we do not consider it.

%%%%%%%%%%
\subsubsection*{$WWZh$ amplitude}
%%%%%%%%%%
In the presence of the Higgs in the spectrum, we have amplitudes with external Higgs particles,
and the cancellation of the $\mathcal{O}(E^2)$ term in these processes puts additional relations among the couplings.
Let us begin with the $WWZh$ amplitude and label $W^+,W^-,Z, h$ as $1,2,3,4$.
The amplitude has the poles at
\begin{align}
	\hat{p}_{12}^2 = m_Z^2,
	\quad
	\hat{p}_{13}^2 = m_W^2,
	\quad
	\hat{p}_{14}^2 = m_W^2,
\end{align}
and is given by
\begin{align}
	A_{WWZh}^{(\lambda_1 \lambda_2 \lambda_3)}
	&= \frac{1}{p_{12}^2-m_Z^2}
	\frac{1}{{z^+_{12} - z^-_{12}}}
	\sum_{\lambda}
	\left[
	{z^+_{12}} \hat{A}_{WWZ}^{(\lambda_1\lambda_2 \lambda )}(z^-_{12}) 
	\times \hat{A}_{ZZh}^{(\bar{\lambda}\lambda_3)}(z^-_{12})
	- {z^-_{12}} \hat{A}_{WWZ}^{(\lambda_1\lambda_2 \lambda)}(z^+_{12}) 
	\times \hat{A}_{ZZh}^{(\bar{\lambda}\lambda_3)}(z^+_{12})
	\right]
	\nonumber \\
	&+ \frac{1}{p_{13}^2-m_W^2}
	\frac{1}{{z^+_{13} - z^-_{13}}}
	\sum_{\lambda}
	\left[
	{z^+_{13}} \hat{A}_{WWZ}^{(\lambda_1\lambda \lambda_3 )}(z^-_{13}) 
	\times \hat{A}_{WWh}^{(\bar{\lambda}\lambda_2)}(z^-_{13})
	- {z^-_{13}} \hat{A}_{WWZ}^{(\lambda_1\lambda\lambda_3)}(z^+_{13}) 
	\times \hat{A}_{WWh}^{(\bar{\lambda}\lambda_2 )}(z^+_{13})
	\right]
	\nonumber \\
	&+\frac{1}{p_{14}^2-m_W^2}
	\frac{1}{{z^+_{14} - z^-_{14}}}
	\sum_{\lambda}
	\left[
	{z^+_{14}} \hat{A}_{WWZ}^{(\lambda \lambda_2 \lambda_3 )}(z^-_{14}) 
	\times \hat{A}_{WWh}^{(\lambda_1\bar{\lambda})}(z^-_{14})
	- {z^-_{14}} \hat{A}_{WWZ}^{(\lambda\lambda_2\lambda_3)}(z^+_{14}) 
	\times \hat{A}_{WWh}^{(\lambda_1\bar{\lambda} )}(z^+_{14})
	\right],
	\label{eq:pole_WWZh}
\end{align}

As usual, we simplify this by using the completeness relation of the intermediate polarization vectors.
With the help of Eq.~\eqref{eq:VVV_Ward}, the products of the three-point amplitudes are given by
\begin{align}
	\hat{A}_{WWZ}^{(\lambda_1\lambda_2 \lambda )}
	\times \hat{A}_{ZZh}^{(\bar{\lambda}\lambda_2)}
	&= -g_{WWZ}g_{ZZh} \epsilon_1^{\dot{a}a}\epsilon_2^{\dot{b}b}\epsilon_3^{\dot{c}c}
	V_{a\dot{a};b\dot{b};c\dot{c}}(\hat{p}_1,\hat{p}_2,-\hat{p}_{12}),
	\\
	\hat{A}_{WWZ}^{(\lambda_1\lambda \lambda_3 )}
	\times \hat{A}_{WWh}^{(\bar{\lambda}\lambda_2)}
	&=  -g_{WWZ}g_{WWh} 
	\left[\epsilon_1^{\dot{a}a}\epsilon_2^{\dot{b}b}\epsilon_3^{\dot{c}c}
	V_{a\dot{a};b\dot{b};c\dot{c}}(\hat{p}_1,-\hat{p}_{13},\hat{p}_{3})
	+ \frac{m_Z^2-m_W^2}{2m_W^2}(\epsilon_1\cdot \epsilon_3)(\epsilon_2 \cdot \hat{p}_{4})\right],
	\\
	\hat{A}_{WWZ}^{(\lambda \lambda_2 \lambda_3 )}
	\times \hat{A}_{WWh}^{(\lambda_1\bar{\lambda})}
	&=  -g_{WWZ}g_{WWh} 
	\left[\epsilon_1^{\dot{a}a}\epsilon_2^{\dot{b}b}\epsilon_3^{\dot{c}c}
	V_{a\dot{a};b\dot{b};c\dot{c}}(-\hat{p}_{23},\hat{p}_{2},\hat{p}_{3})
	- \frac{m_Z^2-m_W^2}{2m_W^2}(\epsilon_2\cdot \epsilon_3)(\epsilon_1 \cdot \hat{p}_{4})\right].
\end{align}
All these terms contain only up to linear in $z_{1i}^\pm$,
and the terms linear in $z_{1i}^\pm$ cancel with each other in Eq.~\eqref{eq:pole_WWZh}.
Therefore the amplitude reduces to the products of the three-point amplitudes without the momentum shift as
\begin{align}
	A_{WWZh}^{(\lambda_1 \lambda_2 \lambda_3)}
	=&
	-\frac{g_{WWZ}g_{ZZh}}{p_{12}^2 - m_Z^2} 
	\epsilon_1^{\dot{a}a}\epsilon_2^{\dot{b}b}\epsilon_3^{\dot{c}c}
	V_{a\dot{a};b\dot{b};c\dot{c}}({p}_1,{p}_2,-{p}_{12})
	\nonumber \\
	&-\frac{g_{WWZ}g_{WWh}}{p_{13}^2 - m_W^2}
	\epsilon_1^{\dot{a}{a}}\epsilon_3^{\dot{c}c}{{V_{a\dot{a};}}^{b\dot{b};}}_{c\dot{c}}({p}_1,-{p}_{13},{p}_{3})
	\left[\epsilon_{bd}\epsilon_{\dot{b}\dot{d}} - \frac{[p_{13}]_{b\dot{b}}[p_{13}]_{d\dot{d}}}{2m_W^2}
	\right]\epsilon_2^{\dot{d}d}
	\nonumber \\
	&-\frac{g_{WWZ}g_{WWh}}{p_{14}^2 - m_W^2}
	\epsilon_2^{\dot{b}{b}}\epsilon_3^{\dot{c}c}{V^{a\dot{a};}}_{b\dot{b};c\dot{c}}(-{p}_{23},{p}_{2},{p}_{3})
	\left[\epsilon_{ad}\epsilon_{\dot{a}\dot{d}} - \frac{[p_{14}]_{a\dot{a}}[p_{14}]_{d\dot{d}}}{2m_W^2}
	\right]\epsilon_1^{\dot{d}d}.
	\label{eq:WWZh_higgs_doublet}
\end{align}
In particular, it does not contain any contact terms.
Its explicit form is given by
%%%%%%%%%%
\begin{align}
	A_{WWZh}^{(\lambda_1 \lambda_2 \lambda_3)}
	&= \frac{g_{ZZh}g_{WWZ} }{m_W^2m_Z(p_{12}^2 - m_Z^2)}
	\bigg[ \langle \textbf{12} \rangle [{\textbf{21}}] \langle {\textbf{3}} \lvert ({p}_1 - {p}_2) \lvert {\textbf{3}} ] 
	- 2 \langle {\textbf{13}} \rangle [{\textbf{31}}] \langle {\textbf{2}} \lvert {p}_1 \lvert {\textbf{2}} ] 
	+ 2  \langle {\textbf{23}} \rangle [{\textbf{32}}] \langle {\textbf{1}} \lvert {p}_2 \lvert {\textbf{1}} ] \bigg]
	\nonumber \\
	&+ \frac{g_{WWZ}g_{WWh} }{m_W^2 m_Z(p_{13}^2 - m_W^2)}
	\nonumber \\
	&\times \left[ \langle {\textbf{13}} \rangle [{\textbf{31}}] 
	\left(\langle {\textbf{2}} \lvert ({p}_1 - {p}_3) \lvert {\textbf{2}} ] 
	- \frac{m_W^2 -m_Z^2}{m_W^2}
	\langle {\textbf{2}} \lvert {p}_4 \lvert {\textbf{2}}]
	\right) 
	- 2 \langle {\textbf{12}} \rangle [{\textbf{21}}] \langle {\textbf{3}} \lvert {p}_1 \lvert {\textbf{3}} ]  
	+ 2  \langle {\textbf{23}} \rangle [{\textbf{32}}] \langle {\textbf{1}} \lvert {p}_3 \lvert {\textbf{1}} ]
	\right]
	\nonumber \\
	&+ \frac{g_{WWZ}g_{WWh} }{m_W^2 m_Z(p_{14}^2 - m_W^2)}
	\nonumber \\
	&\times \left[
	\langle {\textbf{23}} \rangle [{\textbf{32}}] \left(\langle {\textbf{1}} \lvert ({p}_2 - {p}_3) \lvert {\textbf{1}}]
	- \frac{m_Z^2 -m_W^2}{m_W^2} \langle{\textbf{1}} \lvert {p}_4 \lvert {\textbf{1}}]
	\right) 
	- 2 \langle {\textbf{12}} \rangle [{\textbf{21}}] \langle {\textbf{3}} \lvert {p}_2 \lvert {\textbf{3}} ] 
	+ 2  \langle {\textbf{13}} \rangle [{\textbf{31}}] \langle {\textbf{2}} \lvert {p}_3 \lvert {\textbf{2}} ] \right].
\end{align}
The all-longitudinal amplitude behaves as
\begin{align}
	A_{WWZh}^{(LLL)}
	&= \frac{g_{WWZ}(t-u)}{4m_W^4 m_Z}\left(g_{ZZh}m_W^2 - g_{WWh}m_Z^2\right)\ .
\end{align}
At high energy, the cancellation of the $\mathcal{O}(E^2)$ term gives 
\begin{align}
	g_{ZZh}m_W^2 - g_{WWh}m_Z^2 = 0.
\label{eq:constraint3}
\end{align}
The physical meaning of the above equation is straightforward; to unitarize the $WWZh$ amplitudes with only one neutral Higgs that couples to $WW$ and $ZZ$, this Higgs field needs to be fully in charge of the electroweak symmetry breaking, such that the mass ratio between $W$ and $Z$ bosons is their Higgs couplings ratio.

%%%%%%%%%%
\subsubsection*{$WWhh$ amplitude}
%%%%%%%%%%

Now we come to calculate the $WWhh$ scattering.
As we will see, the on-shell method with the ALT shift generates a contact term, 
which guarantees the absence of the $\mathcal{O}(E^2)$ term in this channel.
Therefore, this process does not require any additional relations among the couplings.

Let us label $W^+,W^-,h,h$ as $1,2,3,4$.
The amplitude has the poles at
\begin{align}
	\hat{p}_{13}^2 = m_W^2,
	\quad
	\hat{p}_{14}^2 = m_W^2,
\end{align}
and is given by
\begin{align}
	A_{WWhh}^{(\lambda_1 \lambda_2)}
	&= \sum_{i=3,4}\frac{1}{p_{1i}^2-m_W^2}
	\frac{1}{{z^+_{1i} - z^-_{1i}}}
	\nonumber \\
	&\times\sum_{\lambda}
	\left[
	{z^+_{1i}} \hat{A}_{WWh}^{(\lambda_1 \lambda)}(z^-_{1i}) 
	\times \hat{A}_{WWh}^{(\bar{\lambda}\lambda_2)}(z^-_{1i})
	- {z^-_{1i}} \hat{A}_{WWh}^{(\lambda_1 \lambda)}(z^+_{1i}) 
	\times \hat{A}_{WWh}^{(\bar{\lambda} \lambda_2)}(z^+_{1i})
	\right],
	\label{eq:WWhh_doublet_pole}
\end{align}
where, again, by taking $\lambda_1 = \pm$ we have $B_\infty=0$.
By using the completeness condition of the intermediate polarization vectors,
the product of the three-point amplitude is given by
\begin{align}
	\sum_{\lambda}
	\hat{A}_{WWh}^{(\lambda_1 \lambda)}
	\times \hat{A}_{WWh}^{(\bar{\lambda}\lambda_2)}
	&= -\frac{g_{WWh}^2}{2} \epsilon_1^{\dot{a}{a}} \left[\epsilon_{ab}\epsilon_{\dot{a}\dot{b}}
	-\frac{[\hat{p}_{1i}]_{a\dot{a}}[\hat{p}_{1i}]_{b\dot{b}}}{2m_W^2}
	\right]\epsilon_2^{\dot{b}{b}}.
\end{align}
This contains the $z^2$-term, arising from $\hat{p}_{1i}\hat{p}_{1i}/m_W^2$ in the bracket, 
which results in the contact term.
After simplifying the overall $z$-dependence of the $z^2$-term as before, we obtain
\begin{align}
	A_{WWhh}^{(\lambda_1 \lambda_2)}
	&= 
	-\frac{g_{WWh}^2}{2}
	\sum_{i=3,4}\frac{1}{p_{1i}^2-m_W^2} 
	\epsilon_1^{\dot{a}{a}} \left[\epsilon_{ab}\epsilon_{\dot{a}\dot{b}}
	-\frac{[{p}_{1i}]_{a\dot{a}}[{p}_{1i}]_{b\dot{b}}}{2m_W^2}
	\right]\epsilon_2^{\dot{b}{b}}
	-\frac{g_{WWh}^2}{2m_W^2}\sum_{i=3,4}\frac{(\epsilon_1 \cdot q_{1i})(\epsilon_2 \cdot q_{1i})}{q_1 \cdot q_i},
	\label{eq:WWhh_bfr_simplificaiton_doublet}
\end{align}
where the first term comes from the $z$-independent term, while the second term comes from the $z^2$-term.
Again, the terms linear in $z_{1i}^\pm$ do not contribute since the $z_{1i}^\pm$ contributions cancel with each other.
Recalling that $\lambda_1 = \pm$ and hence $\epsilon_1 = q_1/c_1 m_W$, we can simplify the second term
and obtain
\begin{align}
	A_{WWhh}^{(\lambda_1 \lambda_2)}
	&= -\frac{g_{WWh}^2}{2}
	\sum_{i=3,4}\frac{1}{p_{1i}^2-m_W^2} 
	\epsilon_1^{\dot{a}{a}} \left[\epsilon_{ab}\epsilon_{\dot{a}\dot{b}}
	-\frac{[{p}_{1i}]_{a\dot{a}}[{p}_{1i}]_{b\dot{b}}}{2m_W^2}
	\right]\epsilon_2^{\dot{b}{b}}
	-\frac{g_{WWh}^2}{2m_W^2}\epsilon_1 \cdot \epsilon_2,
	\label{eq:WWhh_final_doublet}
\end{align}
where we used $\sum_i q_i = 0$ and $\epsilon_i \cdot q_i = 0$.
Its explicit form is given by
\begin{align}
	A_{WWhh}^{(\lambda_1 \lambda_2)}
	&= -\frac{g_{WWh}^2}{m_W^2}\sum_{i=3,4}
	\frac{1}{p_{1i}^2-m_W^2}
	\left[\langle \mathbf{12} \rangle [\mathbf{21}]
	+ \frac{\langle \mathbf{1}\vert p_i \vert \mathbf{1}]\langle \mathbf{2}\vert p_j \vert \mathbf{2}]}{m_W^2}
	\right]
	- \frac{g_{WWh}^2}{m_W^4}\langle \mathbf{12} \rangle [\mathbf{21}],
\end{align}
where $j \neq 1,2,i$.
We can see that, with the contact term we derived, $A_{WWhh}^{(LL)}$ behaves only as $\mathcal{O}(E^0)$
at high energy. Therefore, this process does not provide any relations among the couplings.
%%%%%%%%%%
\subsubsection*{$WWtt$ amplitude}
%%%%%%%%%%
After the addition of a single scalar particle, we can write down the following minimal three-point amplitude between scalar and fermions
%%%%%%%%%%%%%%%%%%%%%%%%%%%%%%%%%%%%%%%%
\begin{align}
	A_{ffh}^{(\lambda_1\lambda_2)}
	&= 
	\begin{tikzpicture}[baseline=(v)]
	\begin{feynman}[inline = (base.a), horizontal=p1 to v]
		\vertex (p1) [label=\({\scriptstyle t}\)];
		\vertex [below right = of p1] (v);
		\vertex [below left = of v,label=270:\({\scriptstyle \bar{t}}\)] (p2);
		\vertex [right = of v, label=\({\scriptstyle h}\)] (p3);
		\node [left = -0.25cm of v, blob, fill=gray] (vb);
		\begin{pgfonlayer}{bg}
		\diagram*{
		(p1) -- [fermion,momentum=\({\scriptstyle p_1}\)] (v) -- [fermion,rmomentum'=\({\scriptstyle p_2}\)] (p2),
		(v) -- [scalar, rmomentum'=\({\scriptstyle p_3}\)] (p3),
		};
		\end{pgfonlayer}{bg}
	\end{feynman}
	\end{tikzpicture}
	= g_{ffh}  \left(\langle \textbf{2}\textbf{1} \rangle +[\textbf{2}\textbf{1}]
	\right)
	\end{align}
 %%%%%%%%%%%%%%%%%%%%%%%%%%%%%%
where $g_{ffh}$ is arbitrary dimensionless coupling. Given we have only introduced two quarks $t$ and $b$, $g_{ffh}$ could be $g_{tth}$ or $g_{bbh}$. Then, the Higgs contribution for $WWtt$ amplitude is given by
%%%%%%%%%%%%%%%%%%%%%%%%%%%%%%%%
\begin{align}
	\left.A_{WWtt}^{(\lambda_1 \lambda_2 \lambda_3\lambda_4)}\right\vert_h
	&= \frac{1}{p_{12}^2-m_h^2}
	\frac{1}{{z^+_{12} - z^-_{12}}}
	\nonumber \\
	&\times
	\left[
	{z^+_{12}} \hat{A}_{WWh}^{(\lambda_1 \lambda_2)}(z^-_{12}) 
	\times \hat{A}_{tth}^{(\lambda_3 \lambda_4)}(z^-_{12})
	- {z^-_{12}} \hat{A}_{WWh}^{(\lambda_1 \lambda_2)}(z^+_{12}) 
	\times \hat{A}_{tth}^{(\lambda_3 \lambda_4)}(z^+_{12})
	\right].
	\label{eq:pole_WWtt_higgs}
\end{align}
%%%%%%%%%%%%%%%%%%%%%%%%%%%%%%%%%%
where $z_{12}^{\pm}$ is found by solving $\hat{p}_{12}^2 = m_h^2$. After summing over the pole contributions, we find
%%%%%%%%%%%%%%%%%%%%%%%%%%%%%%%5
\begin{align}
	\left.A_{WWtt}^{(\lambda_1 \lambda_2 \lambda_3\lambda_4)}\right\vert_h
	&= \frac{g_{tth}g_{WWh}}{m_W^2}\frac{(\langle \textbf{1}\textbf{2}\rangle[\textbf{2}\textbf{1}] \langle \textbf{4}\textbf{3} \rangle + \langle \textbf{1}\textbf{2}\rangle[\textbf{2}\textbf{1}] [ \textbf{4}\textbf{3}]) }{p_{12}- m_h^2}
 \label{eq:WWtt_higgs_cont}
 \end{align}
 %%%%%%%%%%%%%%%%%%%%%%%%%%%%%%%%%
Now we take the high energy limit of $WWtt$ amplitude. As in the case of gauge boson amplitude, this amplitude exhibits bad high-energy behavior. $A_{WWtt}^{(LL\pm\mp)}$ scales as $\mathcal{O}(E^2)$ and it vanishes after we impose the following condition among the couplings
\begin{align}
     g_{WWZ}g_{ttZ}^R &= -  g_{WW\gamma}g_{ tt\gamma}, \\
    g_{WWZ}g_{ttZ}^L &= -  g_{WW\gamma}g_{ tt\gamma} +  g_{tbW}^2.
\end{align}
These relations make the leading order energy dependence to be a constant for $A_{WWtt}^{(LL\pm\mp)}$. But, the above relation does not cure the $\mathcal{O}(E)$ scaling of $A_{WWtt}^{(LL\pm\pm)}$ amplitude without the Higgs contribution and unitarity requirement imposes the following constraint among the couplings
% %%%%%%%%%%%%%%%%%%%%%%%%%%%%%%%%%%%%%%%%%%%%%%
 \begin{align}
    {m_t}    g_{tbW}^2  =- g_{tth} g_{WWh} 
    \label{eq:constrant_wwtt}
\end{align}
%%%%%%%%%%%%%%%%%%%%%%

%%%%%%%%%%
\subsubsection*{Coupling relations and $\rho$ parameter}
%%%%%%%%%%

By requiring the cancellation of the $\mathcal{O}(E^2)$ terms, we have obtained three relations in
Eqs.~\eqref{eq:constraint1},~\eqref{eq:constraint2} and~\eqref{eq:constraint3} and they restrict three of the six independent parameters $m_W, m_Z, g_{WWZ}, g_{WW\gamma}, g_{WWh},$ $g_{ZZh}$. 
Two relations fix the Higgs couplings to the gauge bosons as
\begin{align}
	g_{WWh} = g m_W,
	\quad
	g_{ZZh} = \frac{g m_Z^2}{m_W}.
\end{align}
The remaining relation fixes the ratio of the gauge boson masses times the couplings,
known as the $\rho$ parameter
\begin{align}
	\rho = \frac{g^2 m_W^2}{g_{WWZ}^2 m_Z^2} = 1.
\end{align}
In the Lagrangian formulation, it is well known that $\rho$ is linked to the representation
of the Higgs multiplet, and $\rho = 1$ in particular indicates the doublet.
It is interesting that we can derive the same relation, $\rho = 1$, without assuming that the additional scalar particle $h$
is involved in the multiplet of SU(2).
Our result makes sense since the cancellation of the $\mathcal{O}(E^2)$ term effectively forces $h$ 
to be involved in the SU(2) multiplet, and the only option 
of having one additional neutral scalar particle is to embed $h$ into the doublet.

%%%%%%%%%%%%%%%%%%%%%%%%%%%%%%%%%%%%%%%
\subsection{Neutral and doubly charged scalar: Higgs triplet}
\label{subsec:UV_triplet}
%%%%%%%%%%%%%%%%%%%%%%%%%%%%%%%%%%%%%%%

In Sec.~\ref{subsec:UV_doublet}, 
we have seen that the cancellation of $\mathcal{O}(E^2)$ with a single neutral scalar particle
naturally leads to the UV completion by the Higgs doublet, in particular $\rho = 1$.
It is then interesting to study other combinations of the scalar particles 
and see if we conclude that they are also in SU(2) multiplets.
Therefore, in this subsection, we add a neutral scalar, $h^0$, 
and a doubly charged scalar, $h^{++}$, with masses  $m_0$ and $m_2$, respectively.
In the Lagrangian formulation, we know this corresponds to the UV completion by the complex Higgs triplet,
and our focus is to see if we can recover $\rho = 1/2$, corresponding to the Higgs triplet,
in the bottom-up approach.
It turns out that this example reveals a new feature of the on-shell computation;
the Ward identity of four-point amplitudes, the cancellation of the boundary terms,
require relations among seemingly distinct three-point couplings (from the amplitude construction point of view), as we explain below. Of course, all of these are also guaranteed traditionally by the non-broken $U(1)$ gauge symmetry.

The relevant three-point amplitudes are given by
\begin{align}
	A_{WWh^0}^{(\lambda_1\lambda_2)}
	&= \begin{tikzpicture}[baseline=(v)]
	\begin{feynman}[inline = (base.a), horizontal=p1 to v]
		\vertex (p1) [label=\({\scriptstyle W^\pm}\)];
		\vertex [below right = of p1] (v);
		\vertex [below left = of v,label=270:\({\scriptstyle W^\mp}\)] (p2);
		\vertex [right = of v, label=\({\scriptstyle h^0}\)] (p3);
		\node [left = -0.25cm of v, blob, fill=gray] (vb);
		\begin{pgfonlayer}{bg}
		\diagram*{
		(p1) -- [photon,momentum=\({\scriptstyle p_1}\)] (v) -- [photon,rmomentum'=\({\scriptstyle p_2}\)] (p2),
		(v) -- [scalar, rmomentum'=\({\scriptstyle p_3}\)] (p3),
		};
		\end{pgfonlayer}{bg}
	\end{feynman}
	\end{tikzpicture}
	= \frac{g_{WWh}^{(0)}}{ m_W^2} \langle \textbf{1}\textbf{2} \rangle [\textbf{2}\textbf{1}],
	\quad
	A_{ZZh^0}^{(\lambda_1\lambda_2)}
	= 
	\begin{tikzpicture}[baseline=(v)]
	\begin{feynman}[inline = (base.a), horizontal=p1 to v]
		\vertex (p1) [label=\({\scriptstyle Z}\)];
		\vertex [below right = of p1] (v);
		\vertex [below left = of v,label=270:\({\scriptstyle Z}\)] (p2);
		\vertex [right = of v, label=\({\scriptstyle h^0}\)] (p3);
		\node [left = -0.25cm of v, blob, fill=gray] (vb);
		\begin{pgfonlayer}{bg}
		\diagram*{
		(p1) -- [photon,momentum=\({\scriptstyle p_1}\)] (v) -- [photon,rmomentum'=\({\scriptstyle p_2}\)] (p2),
		(v) -- [scalar, rmomentum'=\({\scriptstyle p_3}\)] (p3),
		};
		\end{pgfonlayer}{bg}
	\end{feynman}
	\end{tikzpicture}
	= \frac{g_{ZZh}^{(0)}}{m_Z^2} \langle \textbf{1}\textbf{2} \rangle [\textbf{2}\textbf{1}],
	\\
	A_{WWh^{\mp\mp}}^{(\lambda_1\lambda_2)}
	&= \begin{tikzpicture}[baseline=(v)]
	\begin{feynman}[inline = (base.a), horizontal=p1 to v]
		\vertex (p1) [label=\({\scriptstyle W^\pm}\)];
		\vertex [below right = of p1] (v);
		\vertex [below left = of v,label=270:\({\scriptstyle W^\pm}\)] (p2);
		\vertex [right = of v, label=\({\scriptstyle h^{\mp\mp}}\)] (p3);
		\node [left = -0.25cm of v, blob, fill=gray] (vb);
		\begin{pgfonlayer}{bg}
		\diagram*{
		(p1) -- [photon,momentum=\({\scriptstyle p_1}\)] (v) -- [photon,rmomentum'=\({\scriptstyle p_2}\)] (p2),
		(v) -- [scalar, rmomentum'=\({\scriptstyle p_3}\)] (p3),
		};
		\end{pgfonlayer}{bg}
	\end{feynman}
	\end{tikzpicture}
	= \frac{g_{WWh}^{(2)}}{ m_W^2} \langle \textbf{1}\textbf{2} \rangle [\textbf{2}\textbf{1}],
\end{align}
for those involving one scalar particle, and
\begin{align}
	A_{h^{++}h^{--}Z}^{(\lambda_3)}
	&= \begin{tikzpicture}[baseline=(v)]
	\begin{feynman}[inline = (base.a), horizontal=p1 to v]
		\vertex (p1) [label=\({\scriptstyle h^{++}}\)];
		\vertex [below right = of p1] (v);
		\vertex [below left = of v,label=270:\({\scriptstyle h^{--}}\)] (p2);
		\vertex [right = of v, label=\({\scriptstyle Z}\)] (p3);
		\node [left = -0.25cm of v, blob, fill=gray] (vb);
		\begin{pgfonlayer}{bg}
		\diagram*{
		(p1) -- [scalar,momentum=\({\scriptstyle p_1}\)] (v) -- [scalar,rmomentum'=\({\scriptstyle p_2}\)] (p2),
		(v) -- [photon, rmomentum'=\({\scriptstyle p_3}\)] (p3),
		};
		\end{pgfonlayer}{bg}
	\end{feynman}
	\end{tikzpicture}
	= \frac{g_{hhZ}^{(2)}}{\sqrt{2}m_Z} \langle \mathbf{3}\vert (p_1 - p_2)\vert \mathbf{3}],
	\\
	A_{h^{++}h^{--}\gamma}^{(\lambda_3)}
	&= \begin{tikzpicture}[baseline=(v)]
	\begin{feynman}[inline = (base.a), horizontal=p1 to v]
		\vertex (p1) [label=\({\scriptstyle h^{++}}\)];
		\vertex [below right = of p1] (v);
		\vertex [below left = of v,label=270:\({\scriptstyle h^{--}}\)] (p2);
		\vertex [right = of v, label=\({\scriptstyle \gamma}\)] (p3);
		\node [left = -0.25cm of v, blob, fill=gray] (vb);
		\begin{pgfonlayer}{bg}
		\diagram*{
		(p1) -- [scalar,momentum=\({\scriptstyle p_1}\)] (v) -- [scalar,rmomentum'=\({\scriptstyle p_2}\)] (p2),
		(v) -- [photon, rmomentum'=\({\scriptstyle p_3}\)] (p3),
		};
		\end{pgfonlayer}{bg}
	\end{feynman}
	\end{tikzpicture}
	= \begin{cases}
	\displaystyle \frac{g_{hh\gamma}^{(2)}}{\sqrt{2}\langle 3\xi_3\rangle} 
	\langle \xi_3 \vert (p_1 - p_2)\vert 3], & \mathrm{for}~~\lambda_3 = +,
	\vspace{1mm}
	\\
	\displaystyle -\frac{g_{hh\gamma}^{(2)}}{\sqrt{2}[ 3\xi_3]} 
	\langle 3 \vert (p_1 - p_2)\vert \xi_3], & \mathrm{for}~~\lambda_3 = -,
	\end{cases}
\end{align}
for those involving two scalar particles.
Note that two distinct scalar particles can couple to a vector particle.
In addition, we have Higgs three-point amplitudes, 
but we again ignore them as they do not generate $\mathcal{O}(E^2)$ contributions.

%%%%%%%%%%
\subsubsection*{$WWWW$ amplitude}
%%%%%%%%%%

In the presence of $h^{++}$, the $WWWW$ amplitude has the poles at
\begin{align}
	\hat{p}_{12}^2 = m_0^2,
	\quad
	\hat{p}_{14}^2 = m_0^2,
	\quad
	\hat{p}_{13}^2 = m_2^2.
\end{align}
The calculation is analogous to Sec.~\ref{subsec:UV_doublet},
and therefore we show only the final result, given by
\begin{align}
	\left.A_{WWWW}^{(\lambda_1\lambda_2\lambda_3\lambda_4)}\right\vert_{h}
	&= \left(\frac{g_{WWh}^{(0)}}{m_W^2}\right)^2
	\left[
	\sum_{i=2,4}\frac{\langle \mathbf{1i}\rangle [\mathbf{i1}]\langle \mathbf{3j}\rangle [\mathbf{j3}]}
	{p_{1i}^2 - m_{0}^2}
	\right]
	+ \left(\frac{g_{WWh}^{(2)}}{m_W^2}\right)^2
	\frac{\langle \mathbf{13}\rangle [\mathbf{31}]\langle \mathbf{24}\rangle [\mathbf{42}]}{p_{13}^2 - m_{2}^2}.
	\label{eq:WWWW_higgs_triplet}
\end{align}
Including this contribution, the high energy limit of the all-longitudinal amplitude is given by
\begin{align}
	A_{WWWW}^{(LLLL)}
	&= -\frac{s+u}{4 m_W^4}
	\left[4g^2m_W^2 - 3 g_{WWZ}^2 m_Z^2-\left(g_{WWh}^{(0)}\right)^2
	+ \left(g_{WWh}^{(2)}\right)^2
	\right] + \mathcal{O}(E^0).
\end{align}
Therefore UV completion of this channel requires
\begin{align}
	0 &= 4g^2m_W^2 - 3 g_{WWZ}^2 m_Z^2-\left(g_{WWh}^{(0)}\right)^2
	+ \left(g_{WWh}^{(2)}\right)^2.
\end{align}
In particular, the coupling relation is affected by the presence of the doubly charged scalar.

%%%%%%%%%%
\subsubsection*{$WWZZ$ amplitude}
%%%%%%%%%%
In this process, only the neutral scalar shows up as the intermediate state, 
and we obtain the same result as Sec.~\ref{subsec:UV_doublet}:
\begin{align}
	\left.A_{WWZZ}^{(\lambda_1 \lambda_2 \lambda_3\lambda_4)}\right\vert_h
	&= \frac{g_{WWh}^{(0)} g_{ZZh}^{(0)}}{m_W^2 m_Z^2}
	\frac{\langle \mathbf{12}\rangle [\mathbf{21}]\langle \mathbf{34}\rangle [\mathbf{43}]}
	{p_{12}^2 - m_{0}^2}.
	\label{eq:WWZZ_higgs_triplet}
\end{align}
Including this contribution, the all-longitudinal amplitude behaves as
\begin{align}
	A_{WWZZ}^{(LLLL)}
	&= \frac{t+u}{4m_W^2 m_Z^2}\left[\frac{g_{WWZ}^2 m_Z^4}{m_W^2} 
	- g_{WWh}^{(0)} g_{ZZh}^{(0)}
	\right]
	+ \mathcal{O}(E^0),
\end{align}
and the cancellation of the $\mathcal{O}(E^2)$ term requires
\begin{align}
	g_{WWh}^{(0)} g_{ZZh}^{(0)} &= \frac{g_{WWZ}^2 m_Z^4}{m_W^2}.
\end{align}
%%

%%%%%%%%%%
\subsubsection*{$WWZh$ amplitude}
%%%%%%%%%%

The doubly charged scalar does not affect the $W^+W^- Z h^0$ scattering, 
Therefore, in the same way as Sec.~\ref{subsec:UV_doublet}, this amplitude is given by 
\begin{align}
	A_{WWZh^0}^{(\lambda_1\lambda_2\lambda_3)} 
	=& -\frac{g_{WWZ}g_{ZZh}^{(0)}}{p_{12}^2 - m_Z^2} 
	\epsilon_1^{\dot{a}a}\epsilon_2^{\dot{b}b}\epsilon_3^{\dot{c}c}
	V_{a\dot{a};b\dot{b};c\dot{c}}(\hat{p}_1,\hat{p}_2,-\hat{p}_{12})
	\nonumber \\
	&-\frac{g_{WWZ}g_{WWh}^{(0)}}{p_{13}^2 - m_W^2}
	\epsilon_1^{\dot{a}{a}}\epsilon_3^{\dot{c}c}{{V_{a\dot{a};}}^{b\dot{b};}}_{c\dot{c}}(\hat{p}_1,-\hat{p}_{13},\hat{p}_{3})
	\left[\epsilon_{bd}\epsilon_{\dot{b}\dot{d}} - \frac{[p_{13}]_{b\dot{b}}[p_{13}]_{d\dot{d}}}{2m_W^2}
	\right]\epsilon_2^{\dot{d}d}
	\nonumber \\
	&-\frac{g_{WWZ}g_{WWh}^{(0)}}{p_{14}^2 - m_W^2}
	\epsilon_2^{\dot{b}{b}}\epsilon_3^{\dot{c}c}{V^{a\dot{a};}}_{b\dot{b};c\dot{c}}(-\hat{p}_{23},\hat{p}_{2},\hat{p}_{3})
	\left[\epsilon_{ad}\epsilon_{\dot{a}\dot{d}} - \frac{[p_{14}]_{a\dot{a}}[p_{14}]_{d\dot{d}}}{2m_W^2}
	\right]\epsilon_1^{\dot{d}d},
\end{align}
and the high-energy limit is given by
\begin{align}
	A_{WWZh^0}^{(\lambda_1\lambda_2\lambda_3)}
	&= \frac{g_{WWZ}(t-u)}{4m_W^4 m_z}\left[g_{ZZh}^{(0)}m_W^2 - g_{WWh}^{(0)}m_Z^2\right]
	+ \mathcal{O}(E^0).
\end{align}
From this result, we obtain the relation
\begin{align}
	0 &= g_{ZZh}^{(0)}m_W^2 - g_{WWh}^{(0)}m_Z^2.
\end{align}
This is the same condition as the doublet case in Eq.~\ref{eq:constraint3} and the interpretation is the same.

In the presence of the doubly charged scalar, there is also the $W^+ W^+ Z h^{--}$ four-point scattering amplitude.
We label $W^+, W^-, Z, h^{--}$ as $1,2,3,4$, respectively.
In this process, we have the poles at
\begin{align}
	\hat{p}_{12}^2 = m_{2}^2,
	\quad
	\hat{p}_{13}^2 = m_W^2,
	\quad
	\hat{p}_{14}^2 = m_W^2.
\end{align}
The products of the three-point amplitudes contain the terms up to only linear in $z_{1i}^\pm$,
and therefore we obtain
\begin{align}
	A_{WWZh^{--}}^{(\lambda_1\lambda_2\lambda_3)}
	=& 
	- \frac{2g_{WWh}^{(2)}g_{hhZ}^{(2)}}{p_{12}^2 -m_{2}^2} (\epsilon_1 \cdot \epsilon_2)
	(\epsilon_3 \cdot p_4)
	\nonumber \\
	&-\sum_{i=1,2}
	\left[
	\frac{g_{WWZ}g_{WWh}^{(2)}}{p_{i3}^2 - m_W^2} \epsilon_i^{\dot{a}{a}} \epsilon_3^{c\dot{c}} 
	{{V_{a\dot{a};}}^{b\dot{b};}}_{c\dot{c}}
	(p_i,-p_{i3},p_3)
	\left(
	\epsilon_{bd}\epsilon_{\dot{b}\dot{d}}
	-\frac{[p_{i3}]_{b\dot{b}}[p_{i3}]_{d\dot{d}}}{2m_W^2}
	\right)
	\epsilon_{j}^{d\dot{d}}
	\right].
	\label{eq:WWZh_triplet}
\end{align}
The high-energy limit of the all-longitudinal amplitude is given by
\begin{align}
	A_{WWZh^{--}}^{(LLL)}
	&= \frac{t+u}{4m_W^4 m_Z}
	\left[
	g_{WWZ} g_{WWh}^{(2)} (4m_W^2 - m_Z^2)
	+ 2m_W^2g_{WWh}^{(2)}g_{hhZ}^{(2)}
	\right]
	+ \mathcal{O}(E^0).
\end{align}
Therefore we obtain
\begin{align}
	0 &= 
	g_{WWZ} g_{WWh}^{(2)} \left(4 - \frac{m_Z^2}{m_W^2}\right)
	+2g_{WWh}^{(2)}g_{hhZ}^{(2)}.
\end{align}
%%

%%%%%%%%%%
\subsubsection*{$WW\gamma h$ amplitude}
%%%%%%%%%%

To fix $g_{hh\gamma}^{(2)}$, we consider the $W^+ W^+ \gamma h^{--}$ scattering. For the external photon, we fix the reference vector so that the transverse polarization vector has vanishing temporal component and the photon momentum is shifted by this transverse polarization vector.
The computation is analogous to $WWZh^{--}$.
The amplitude has the poles at
\begin{align}
	\hat{p}_{12}^2 = m_2^2,
	\quad
	\hat{p}_{13}^2 = m_W^2,
	\quad
	\hat{p}_{14}^2 = m_W^2,
\end{align}
and the pole contributions contain terms only up to linear in $z$.
Therefore we obtain
\begin{align}
	A_{WW\gamma h^{--}}^{(\lambda_1\lambda_2\lambda_3)}
	=& -\frac{2g_{WWh}^{(2)}g_{hh\gamma}^{(2)}}{p_{12}^2 - m_2^2}
	(\epsilon_1 \cdot \epsilon_2) (\epsilon_3 \cdot p_4)
	\nonumber \\
	&-\sum_{i=1,2}\frac{g_{WWh}^{(2)}g_{WW\gamma}}{p_{i3}^2-m_W^2}
	\epsilon_i^{\dot{a}a}\epsilon_3^{\dot{c}c}{{V_{a\dot{a};}}^{b\dot{b};}}_{c\dot{c}}(p_i,-p_{i3},p_3)
	\left[\epsilon_{bd}\epsilon_{\dot{b}\dot{d}}-\frac{[p_{i3}]_{b\dot{b}}[p_{i3}]_{d\dot{d}}}{2m_W^2}
	\right]
	\epsilon_j^{\dot{d}d},
	\label{eq:WWgammah_triplet}
\end{align}
where $j\neq 3,4,i$.
Now, if we replace the photon polarization vector by its momentum, $\epsilon_3 \to p_3$, we obtain
\begin{align}
	\left.A_{WW\gamma h^{--}}^{(\lambda_1\lambda_2\lambda_3)}\right\vert_{\epsilon_3 \to p_3}
	=& -\left(g_{hh\gamma}^{(2)}+2g_{WW\gamma}\right)(\epsilon_1\cdot \epsilon_2).
\end{align}
This indicates that the Ward identity of the photon does not hold for arbitrary $g_{hh\gamma}^{(2)}$, 
and we need
\begin{align}
	g_{hh\gamma}^{(2)} = -2g_{WW\gamma}.
\end{align}

We require this condition for the consistency of our procedure from both gauge symmetry perspective and constructibility consistency, i.e., $B_\infty=0$.
This relation makes sense since we assume that $h^{++}$ is doubly charged under the photon,
but the on-shell method does not automatically tell us this relation.
Seemingly, if the gauge symmetry relates different three-point couplings such as $g_{hh\gamma}^{(2)}$
and $g_{WW\gamma}$, we need to impose these conditions.

%%%%%%%%%%
\subsubsection*{$WWhh$ amplitude} 
%%%%%%%%%%
\label{sec:WWhh_triplet}
Finally, we consider the $WWhh$ scattering. The $W^+ W^- h^0 h^0$ amplitude is not affected by $h^{++}$,
while $W^+ W^+ h^0 h^{--}$ is analogous to the $W^+ W^- h^0 h^0$. Therefore, we focus on the amplitude of $W^+ W^- h^{++} h^{--}$, labeled as $1,2,3,4$, respectively.
In this process, we have the poles at
\begin{align}
	\hat{p}_{12}^2 = m_Z^2,\,\mathrm{or}\,~0,
	\quad
	\hat{p}_{14}^2 = m_W^2,
\end{align}
and the amplitude is given by
\begin{align}
	A_{WWh^{++}h^{--}}^{(\lambda_1\lambda_2)}
	=& \sum_{I=Z,\gamma}\frac{1}{p_{12}^2-m_I^2}
	\frac{1}{z_{12}^+-z_{12}^-}
	\sum_\lambda\left[z_{12}^+ \hat{A}_{WWI}^{(\lambda_1\lambda_2\lambda)}(z_{12}^-)
	\times \hat{A}_{h^{++}h^{--}I}^{(\bar{\lambda})}(z_{12}^-) - (z_{12}^+ \leftrightarrow z_{12}^-)
	\right]
	\nonumber \\
	&+ \frac{1}{p_{14}^2 - m_W^2}
	\frac{1}{z_{14}^+ - z_{14}^-}
	\sum_\lambda\left[z_{14}^+\hat{A}_{WWh^{--}}^{(\lambda_1\lambda)}(z_{14}^-)
	\times \hat{A}_{WWh^{++}}^{(\bar{\lambda}\lambda_2)}(z_{14}^-)
	- (z_{14}^+\leftrightarrow z_{14}^-)
	\right],
\end{align}
where, as usual, we assume $\lambda_1 = \pm$.
The products of the three-point amplitudes are given by
\begin{align}
	\hat{A}_{WWI}^{(\lambda_1\lambda_2\lambda)}
	\times \hat{A}_{h^{++}h^{--}I}^{(\bar{\lambda})}
	&= -g_{WWI}g_{hhI}^{(2)}\epsilon_1^{\dot{a}a}\epsilon_2^{\dot{b}b}
	V_{a\dot{a};b\dot{b};c\dot{c}}(\hat{p}_1,\hat{p}_2,-\hat{p}_{12}) (\hat{p}_3-\hat{p}_4)^{\dot{c}c},
	\\
	\hat{A}_{WWh^{--}}^{(\lambda_1\lambda)}
	\times \hat{A}_{WWh^{++}}^{(\bar{\lambda}\lambda_2)}
	&= -\frac{\left(g_{WWh}^{(2)}\right)^2}{2}\epsilon_1^{\dot{a}a}\left[
	\epsilon_{ab}\epsilon_{\dot{a}\dot{b}} - \frac{[\hat{p}_{14}]_{a\dot{a}}[\hat{p}_{14}]_{b\dot{b}}}{2m_W^2}
	\right]\epsilon_2^{b\dot{b}}
\end{align}
Each term contains up to the $z^2$-terms.
By using $\lambda_1 = \pm$ and hence $q_1 = c_1 m_W \epsilon_1$,
the $z^2$-term is given by
\begin{align}
	\left[A_{WWh^{++}h^{--}}^{(\lambda_1\lambda_2)}\right]_\mathrm{contact}
	&= \frac{1}{\sqrt{2} c_1 m_W}
	\left[
	\left(g_{WWZ}g_{hhZ}^{(2)} + g_{WW\gamma}g_{hh\gamma}^{(2)}\right) 
	\langle \mathbf{2}\vert (q_3 - q_4)\vert \mathbf{2}]
	+ \frac{\left(g_{WWh}^{(2)}\right)^2}{2m_W^2} \langle \mathbf{2}\vert q_3 \vert \mathbf{2}]
	\right].
	\label{eq:WWhh_triplet_bfr_cond}
\end{align}
For arbitrary choice of $g_{hhZ}^{(2)}$ and $g_{WWh}^{(2)}$, this \emph{does not} lead to the result
independent of $q_i$.
Since the original amplitude does not depend on $q_i$ (except for $q_1 \propto \epsilon_1$), 
we need a non-zero $B_\infty$ to cancel the $q_i$-dependence in general, 
indicating that the Ward identity does not hold for arbitrary $g_{hhZ}^{(2)}$ and $g_{WWh}^{(2)}$.
In the current case, the result is independent of these variables only when we have
\begin{align}
	0 &= 4\left(g_{WWZ}g_{hhZ}^{(2)} + g_{WW\gamma} g_{hh\gamma}^{(2)}\right) m_W^2
	+ \left(g_{WWh}^{(2)}\right)^2,
 \label{eq:WWhh_triplet}
\end{align}
and we require this relation in the following, for the consistency of seeking theories constructible within the ALT shift.
With this condition, we obtain the contact term as
\begin{align}
	\left[A_{WWh^{++}h^{--}}^{(\lambda_1\lambda_2)}\right]_\mathrm{contact}
	&= -\frac{\left(g_{WWh}^{(2)}\right)^2}{4m_W^4}\langle \mathbf{12}\rangle [\mathbf{21}].
	\label{eq:WWhh_triplet_contact}
\end{align}
Again, we learn the same lesson as the $WW\gamma h^{--}$ scattering;
the on-shell method automatically reproduces the relations between three- and four-point amplitudes
as we have seen throughout this paper,
but if the gauge symmetry relates three-point couplings that are seemingly different
from the bottom-up point of view, we need to impose them by hand
in the on-shell method.
%%%%%%%%%%
\subsubsection*{$WWtt$ amplitude}
%%%%%%%%%%
We obtain similar relation as in Eq.~\eqref{eq:constrant_wwtt} for this case since there is no additional $tth^{++}$ or $tth^{--}$ three-point amplitude. We find the relation
 %%%%%%%%%%%%%%%%%%%%%
 \begin{align}
    {m_t}   g_{tbW}^2  = -g_{htt}^{(0)} g_{WWh}^{(0)} 
    \label{eq:constrant_triplet}
\end{align}
%%%%%%%%%%%%%%%%%%%%%%
%%%%%%%%%%
\subsubsection*{Coupling relations and $\rho$ parameter}
%%%%%%%%%%

By requiring the cancellations of the $\mathcal{O}(E^2)$ term, we have obtained
\begin{align}
	0 &= 4g^2m_W^2 - 3 g_{WWZ}^2 m_Z^2-\left(g_{WWh}^{(0)}\right)^2
	+ \left(g_{WWh}^{(2)}\right)^2,
	\\
	0 &= g_{WWh}^{(0)} g_{ZZh}^{(0)} - \frac{g_{WWZ}^2 m_Z^4}{m_W^2},
	\\
	0 &= g_{ZZh}^{(0)}m_W^2 - g_{WWh}^{(0)}m_Z^2,
	\\
	0 &= 
	g_{WWZ} g_{WWh}^{(2)} \left(4 - \frac{m_Z^2}{m_W^2}\right)
	+2g_{WWh}^{(2)}g_{hhZ}^{(2)}.
\end{align}
Moreover, we have seen that we need to impose additional relations among three-point couplings 
to satisfy the Ward identity of the four-point amplitudes. These are given by
\begin{align}
	0 &= g_{hh\gamma}^{(2)} + 2g_{WW\gamma},
	\\
	0 &= 4\left(g_{WWZ}g_{hhZ}^{(2)} + g_{WW\gamma} g_{hh\gamma}^{(2)}\right) m_W^2
	+ \left(g_{WWh}^{(2)}\right)^2.
\end{align}
These relations fix the couplings between the Higgs and the gauge bosons as
\begin{align}
	g_{WWh}^{(0)} = g_{WWZ}m_Z,
	\quad
	g_{ZZh}^{(0)} = \frac{g_{WWZ}m_Z^3}{m_W^2},
	\quad
	\left(g_{WWh}^{(2)}\right)^2 = 4m_W^2
	\left[g_{WWZ}^2\left(2-\frac{m_Z^2}{2m_W^2}\right)+2g_{WW\gamma}^2\right],
\end{align}
and
\begin{align}
	g_{hh\gamma}^{(2)} = -2g_{WW\gamma},
	\quad
	g_{hhZ}^{(2)} = -g_{WWZ}\left(2-\frac{m_Z^2}{2m_W^2}\right).
\end{align}
In addition, the $\rho$ parameter is fixed as
\begin{align}
	\rho = \frac{g^2 m_W^2}{g_{WWZ}^2 m_Z^2} = \frac{1}{2}.
\end{align}
Therefore, our calculation suggests that the doubly charged and neutral scalars
for this unitarization are in the SU(2) triplet. 

%%%%%%%%%%%%%%%%%%%%%%%%%%%%%%%%%%%%%%%
\subsection{Other UV completion}
%%%%%%%%%%%%%%%%%%%%%%%%%%%%%%%%%%%%%%%

So far we have studied UV completion of the EW theory by adding scalar particles.
However, UV completion by scalar particles is not the only option (forgetting about the experimental results) and we can add e.g.~vector particles to improve
the high-energy behavior of the amplitudes.
For instance, we can add a vector particle, $Z'$, which couples to the $W$-boson with the coupling $g_{WWZ'}$.
Then, in the same way as Sec.~\ref{subsec:EW_WWWW}, we obtain the high-energy behavior of the $WWWW$ scattering as
\begin{align}
	\left.A_{WWWW}^{(LLLL)}\right\vert_{Z,\gamma}
	&= -\frac{s+u}{4 m_W^4}
	(4g^2m_W^2 - 3 g_{WWZ}^2 m_Z^2 - 3g_{WWZ'}^2m_{Z'}^2) + \mathcal{O}(E^0),
\end{align}
where $m_{Z'}$ is the mass of the $Z'$-boson,
and we now have $g^2 = g_{WWZ}^2 + g_{WW\gamma}^2 + g_{WWZ'}^2$.
Therefore $Z'$ can unitarize this specific channel if
\begin{align}
	0 &= 4g^2m_W^2 - 3 g_{WWZ}^2 m_Z^2 - 3g_{WWZ'}^2m_{Z'}^2.
\end{align}
In this case, we need to study amplitudes with external $Z'$, 
and it is known that in the end, we need an inifnite tower of spin-1 particles, 
as discussed in~\cite{Csaki:2003dt,Csaki:2003zu}.
Instead, we can try UV completion by higher-spin particles such as spin-2, 3, and so on.
The on-shell method is suitable for this purpose 
since it allows us to write down the amplitudes with arbitrary spins, at least in principle.
We leave a detailed study on possible UV completions of the EW theory,
including higher spin particles, in the on-shell formalism as a future work.

%%%%%%%%%%%%%%%%%%%%%%%%%%%%%%%%%%%%%%%
\section{Conclusion and discussion}
\label{sec:conclusion}
%%%%%%%%%%%%%%%%%%%%%%%%%%%%%%%%%%%%%%%
%%%%%%%%%%%%%%%%%%%%%%%%%%
\begin{figure}[t!] 
  \centering
    \includegraphics[width=0.9\textwidth]{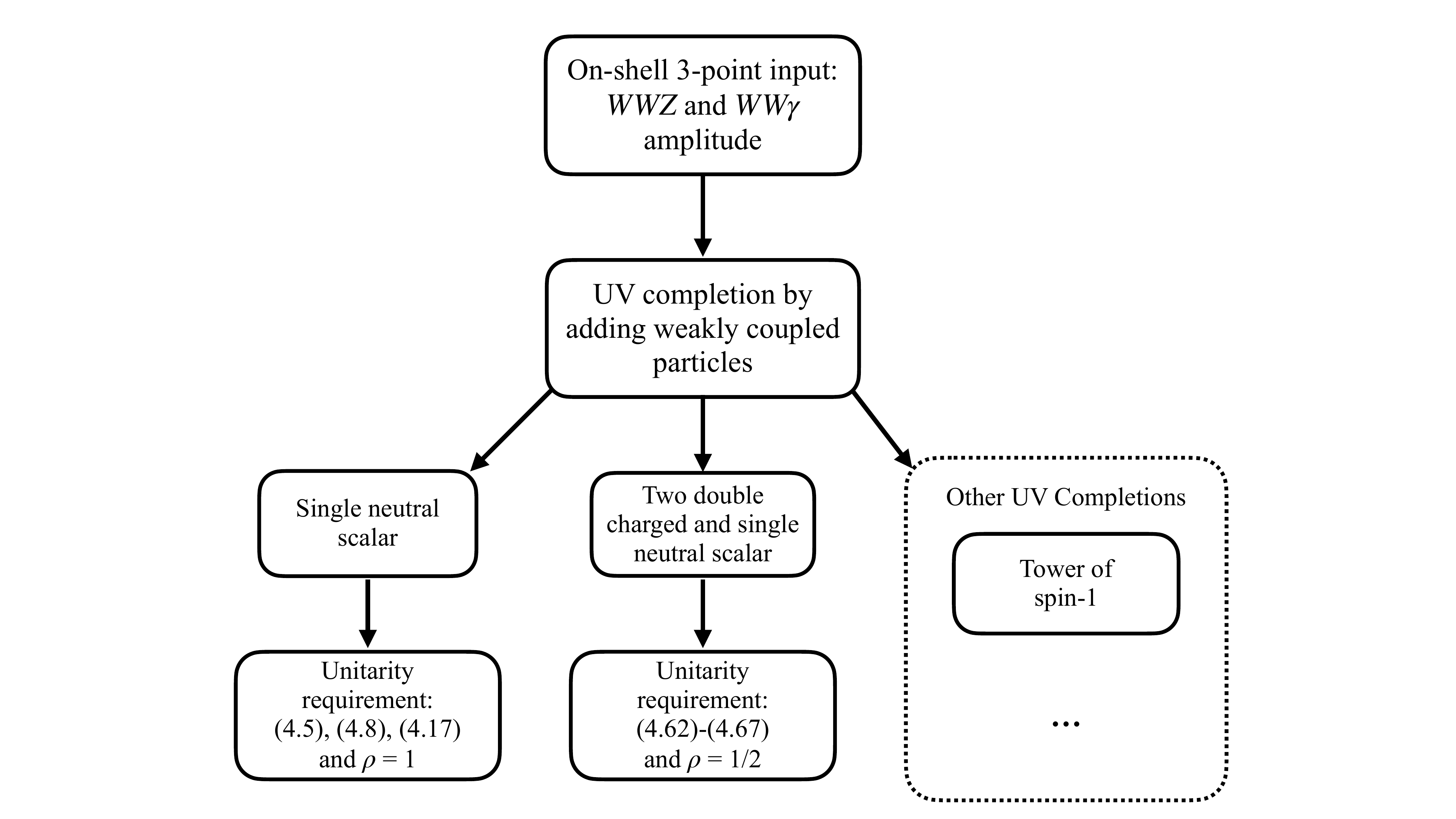}
\caption{Summary of recursive construction of EW theory.}
\label{fig:schematic_flow_2}
\end{figure}
%%%%%%%%%%%%%%%%%%%%%%%%%%%%%%%

On-shell recursion relations are a key component of the amplitude program. They have proven to be very powerful and efficient in constructing $n$-point amplitudes of massless particles, compared to the usual Feynman diagram approach. With the progress of the little-group covariant massive spinor-helicity formalism~\cite{Arkani-Hamed:2017jhn}, a natural direction to pursue is to build massive higher-point amplitudes from lower-point ones. However, without an appropriate momentum shift, such construction suffers from contact term ambiguity because of the non-vanishing boundary terms. 

To address this issue, we have proposed the ALT shift, which is applicable in the same way for massive and massless theories, in~\cite{Ema:2024vww} that generates no boundary term in massive QED and electroweak theories, or more generally renormalizable theories with spin $\leq1$. In addition, we have highlighted in this paper another remarkable feature of the ALT shift, which is its connection to Ward identity. Under the ALT shift and imposing the Ward identity of the massless theory from the beginning, one not only pins down the (spontaneously broken) gauge theory but also guarantees the on-shell constructibility. This connection is closely related to the fact that the shifted momentum is proportional to the transverse polarization vector, which is a feature of the proposed ALT shift. As a consequence, the contact terms coming from gauge symmetry naturally arise under the ALT shift.

Numerous examples of four-point amplitudes are presented in this work. We schematically summarize them in \autoref{fig:schematic_flow_2}. First, without the introduction of the Higgs boson, we have recovered all the contact terms in electroweak theory that cancel the fastest energy growths of the factorizable part. The Standard Model Higgs scalar can then be incorporated by the unitarity requirement. We have also successfully reproduced the Higgs coupling to gauge bosons from the bottom-up using the on-shell constructive method. Beyond the Standard Model, our ALT shift can be applied to other renormalizable UV completions, and we have showcased the one with a Higgs triplet. Here again, we have correctly reproduced the relations between couplings by requiring unitarity and occasionally imposing gauge symmetry via demanding Ward identity. 

The method developed here and in our previous work~\cite{Ema:2024vww} may lead to many interesting applications. Apart from the Higgs doublet and triplet, other UV completions can be studied with the ALT shift in the on-shell framework. It may also be used to study e.g.~EW radiations within the SM at high energy relevant for future colliders
(see e.g.~\cite{Nardi:2024tce} as a recent related work). 
Beyond the context of the EW theory, one may also attempt to generalize ALT shift for massive higher spin amplitude, which can be applied to analyze Compton amplitude of general spinning compact objects \cite{Chung:2018kqs,Chen:2021kxt,Chiodaroli:2021eug,Haddad:2023ylx} as well as study its generalization to higher dimensional massive spinors \cite{Pokraka:2024fao}. The ALT shift may be useful for understanding the general properties of tree-level massive scattering amplitudes.
For instance, the color-kinematic duality and double copy relations of massless gauge theory and gravity are 
proven at tree level with an arbitrary number of external
particles~\cite{Kawai:1985xq,Bern:2008qj,HenryTye:2010tcy,Feng:2010my,Bern:2010yg,
Bjerrum-Bohr:2010pnr,Mafra:2011kj,Fu:2012uy}.
Here the on-shell recursion relation plays a crucial role
as it allows us to study the general properties of the amplitudes based on mathematical induction. 
Therefore, we may envision that the ALT shift will enable us 
to extend our understanding of these properties to full massive amplitudes.

%%%%%%%%%%%%%%%%%%%%%%%%%%%%%%%%%%%%%%%
\section*{Acknowledgements}
%%%%%%%%%%%%%%%%%%%%%%%%%%%%%%%%%%%%%%%

We thank N. Christensen, B. Feng, and Y. Shadmi for helpful discussion.
This work is supported in part by U.S.~Department of Energy Grant No.~DESC0011842.
Y.E.~and Z.L.~acknowledge the Aspen Center for Physics, where the final part of the work was performed, which is supported by National Science Foundation grant PHY-2210452.
The (Feynman-like) diagrams in this paper are drawn with \texttt{TikZ-Feynman}~\cite{Ellis:2016jkw}. 

%%%%%%%%%%%%%%%%%%%%%%%%%%%%%%%%%%%%%%%
%%%%%%%%%%%%%%%%%%%%%%%%%%%%%%%%%%%%%%%
\appendix
%%%%%%%%%%%%%%%%%%%%%%%%%%%%%%%%%%%%%%%
%%%%%%%%%%%%%%%%%%%%%%%%%%%%%%%%%%%%%%%

%%%%%%%%%%%%%%%%%%%%%%%%%%%%%%%%%%%%%%%
\section{Massive spinor formalism}
\label{app:convention}
%%%%%%%%%%%%%%%%%%%%%%%%%%%%%%%%%%%%%%%

In this appendix, we summarize the massive spinor formalism and conventions used in this paper.
We express the four-momentum $p^\mu$ in the spinor space as
\begin{align}
	p_{a\dot{a}} = p_\mu \left[\sigma^\mu\right]_{a\dot{a}} = 
	\begin{pmatrix} p^0 - p^3 & -p^1 + i p^2 \\ -p^1 - ip^2 & p^0 + p^3\end{pmatrix},
	\quad
	p^{\dot{a}a} = p_\mu\left[\bar{\sigma}^\mu\right]^{\dot{a}a} = 
	\begin{pmatrix} p^0 + p^3 & p^1 - i p^2 \\ p^1+ ip^2 & p^0 - p^3\end{pmatrix}.
\end{align}
The determinant is given by
\begin{align}
	\mathrm{det}\left[p_{a\dot{a}}\right] = \mathrm{det}\left[p^{\dot{a}a}\right] = m^2,
\end{align}
where $m$ is the mass of the particle.
We also note that
\begin{align}
	p_i^{\dot{a}a}(p_j)_{a\dot{a}} = 2p_i \cdot p_j,
	\quad
	\epsilon^{ab}\epsilon^{\dot{a}\dot{b}}p_{b\dot{b}} = (p^{\dot{a}a})^T.
\end{align}
%%

%%%%%%%%%%
\subsubsection*{Massive spinor}
%%%%%%%%%%

In general, if $m_i\neq 0$ for a particle $i$, we can write the momentum $p_i$ as~\cite{Arkani-Hamed:2017jhn}
\begin{align}
	[p_{i}]_{a\dot{a}} = \vert i\rangle_{a}^I [i\vert_{\dot{a}I},
	\quad
	I = 1,2.
\end{align}
Correspondingly we have
\begin{align}
	[p_i]^{\dot{a}a} = \vert i]_{I}^{\dot{a}} \langle i\vert^{aI}.
\end{align}
The little group is the transformation that keeps the momentum unchanged.
Therefore in this expression, the little group acts on the indices $I$ and is SU(2) for real momentum.
By taking its determinant we obtain
\begin{align}
	\mathrm{det}[\vert i\rangle_a^I]\times\mathrm{det}[[i\vert_{\dot{a}I}] = m_i^2.
\end{align}
We fix the relative normalization as
\begin{align}
	\mathrm{det}[\vert i\rangle_{a}^I] = \mathrm{det}[[i\vert_{\dot{a}I}] = m_i.
\end{align}
We raise and lower the indices $I$, $J$ 
by the anti-symmetric tensor $\epsilon^{IJ}$ and $\epsilon_{IJ}$ with $\epsilon^{12} = -\epsilon_{12} = 1$ as
\begin{align}
	\lambda^I = \epsilon^{IJ}\lambda_J,
	\quad
	\lambda_I = \epsilon_{IJ}\lambda^J.
\end{align}
Since $\epsilon_{IJ}\vert i\rangle^{I}_a \vert i\rangle^{J}_b$ is anti-symmetric with respect to $a$ and $b$, it should
be proportional to $\epsilon_{ab}$, and the same holds for $\epsilon^{IJ}[i\vert_{\dot{a}I} [i\vert_{\dot{b}J}$.
The determinant then fixes the overall normalization as
\begin{align}
	\epsilon_{IJ}\vert i\rangle^{I}_a \vert i\rangle^{J}_b = m_i \epsilon_{ab},
	\quad
	\epsilon^{IJ}[i\vert_{\dot{a}I}[i\vert_{\dot{b}J} = -m_i\epsilon_{\dot{a}\dot{b}},
	\label{eq:hol_onshell}
\end{align}
and similarly
\begin{align}
	\langle i^I i^J\rangle = -m_i \epsilon^{IJ},
	\quad
	[i_{I} i_{J}] = -m_i\epsilon_{IJ},
	\label{eq:mass_lambda}
\end{align}
where we denote
\begin{align}
	\langle i j\rangle = \langle i \vert^{a} \vert j\rangle_{a}
	= \epsilon^{ab}\vert i\rangle_{b} \vert j\rangle_a,
	\quad
	[ij] = [i\vert_{\dot{a}} \vert j]^{\dot{a}}
	= \epsilon_{\dot{a}\dot{b}}\vert j]^{\dot{a}}\vert i]^{\dot{b}}.
\end{align}
From these relations, we obtain
\begin{align}
	\langle i\vert^{aI} [p_{i}]_{a\dot{a}} = -m_i[i\vert_{\dot{a}}^{I},
	\quad
	[p_{i}]_{a\dot{a}}\vert i]^{\dot{a}I} = m_i\vert i\rangle_{a}^I,
\end{align}
which is nothing but the Dirac equation.
In the same way, we obtain
\begin{align}
	p^{\dot{a}a}\vert i\rangle^{I}_a = m_i\vert i]^{\dot{a}I},
	\quad
	[i\vert_{\dot{a}}^{I}p^{\dot{a}a} = -m_i\langle i\vert^{aI}.
\end{align}
We use the bold notation introduced in~\cite{Arkani-Hamed:2017jhn} 
to express spinors with floating little-group indices, when no ambiguity would arise with the abbreviation.

%%%%%%%%%%
\subsubsection*{Massless spinor}
%%%%%%%%%%

If $m = 0$, we need only one pair of spinors to express the momentum as
\begin{align}
	p_{a\dot{a}} = \vert p\rangle_{a} [ p\vert_{\dot{a}}.
\end{align}
In this case the little group is U(1) for real momentum. For complex momentum, one can extend this to the little-group scaling.

%%%%%%%%%%
\subsubsection*{Dirac wavefunctions}
%%%%%%%%%%

The Dirac wavefunctions, $u$ and $v$, satisfy the Dirac equation
\begin{align}
	\left[\slashed{p}-m\right] u(p) = 0,
	\quad
	\left[\slashed{p}+m\right] v(p) = 0.
\end{align}
One can easily see that this is solved as
\begin{align}
	u^I(p_i) = \begin{pmatrix} \vert i\rangle_a^{I} \\ \vert i]^{\dot{a}I}\end{pmatrix},
	\quad
	v^I(p_i) = \begin{pmatrix} \vert i\rangle_a^{I} \\ -\vert i]^{\dot{a}I}\end{pmatrix}.
\end{align}
Note that $u(-p)$ and $v(p)$ satisfy the same equation, and hence 
\begin{align}
	u^I(-p_i) = \begin{pmatrix} \vert i\rangle_a^{I} \\ -\vert i]^{\dot{a}I}\end{pmatrix},
	\quad
	v^I(-p_i) = \begin{pmatrix} \vert i\rangle_a^{I} \\ \vert i]^{\dot{a}I}\end{pmatrix}.
\end{align}
In other words, we have the relations
\begin{align}
	\vert -i\rangle^I = \vert i\rangle^I,
	\quad
	\vert -i]^I = -\vert i]^I.
	\label{eq:spinor_negative_mom}
\end{align}
In the same way we can show that
\begin{align}
	\bar{u}^I(p) = \begin{pmatrix} -\langle i\vert^{aI} & [i\vert_{\dot{a}}^{I} \end{pmatrix},
	\quad
	\bar{v}^{I}(p) = \begin{pmatrix} \langle i\vert^{aI} & [i\vert_{\dot{a}}^{I}\end{pmatrix}.
\end{align}
They satisfy
\begin{align}
	\epsilon_{IJ}u^I(p)\bar{u}^J(p) = \slashed{p} + m,
	\quad
	\epsilon_{IJ}v^{I}(p)\bar{v}^{J}(p) = \slashed{p}-m. \label{eq:fermion_complt}
\end{align}
%%

%%%%%%%%%%
\subsubsection*{Helicity basis and polarization vector}
%%%%%%%%%%

We work in the helicity basis to define the ALT shift in the main text,
and here we summarize its basic properties.
The solution of the Dirac equation is given by
\begin{align}
	u^I(p) = \begin{pmatrix} \sqrt{p\cdot \sigma} \xi^I \\ \sqrt{p\cdot \bar{\sigma}} \xi^I\end{pmatrix}.
\end{align}
We can choose $\xi^I$ as the eigenstates of the helicity operator $\hat{h}$, given by
\begin{align}
	\hat{h} = \frac{\hat{p}\cdot \vec{\sigma}}{2}
	= \frac{1}{2}\begin{pmatrix} \cos\theta & \sin\theta e^{-i\phi} \\ \sin\theta e^{i\phi} & -\cos\theta\end{pmatrix},
\end{align}
where $(\theta,\phi)$ denotes the solid angle of the momentum.
The eigenvalues are $\pm 1/2$ and the eigenstates are
\begin{align}
	\xi^+ = \begin{pmatrix} c \\ s \end{pmatrix},
	\quad
	\xi^- = \begin{pmatrix} -s^* \\ c \end{pmatrix},
	\quad
	\mathrm{where}
	~~
	c = \cos\frac{\theta}{2},
	~~
	s = e^{i\phi}\sin\frac{\theta}{2},
\end{align}
which satisfy $\hat{h} \xi^\pm = \pm\xi^\pm/2$.
As we saw before, the angle and square brackets are given by
\begin{align}
	\vert p\rangle_{a}^{I} = \sqrt{p\cdot \sigma} \xi^I,
	\quad
	\vert {p}]^{\dot{a}I} = \sqrt{p\cdot \bar{\sigma}} \xi^I.
\end{align}
With this identification, we obtain
\begin{align}
	&\vert {p}\rangle_a^{I} = \sqrt{E-p}\begin{pmatrix} c \\ s \end{pmatrix} \delta^{I}_+ 
	+ \sqrt{E+p}\begin{pmatrix} -s^* \\ c \end{pmatrix} \delta^{I}_-,
	\\
	&[{p}\vert_{\dot{a}}^{I} = \sqrt{E+p}\begin{pmatrix} -s & c \end{pmatrix} \delta^{I}_+
	- \sqrt{E-p}\begin{pmatrix} c & s^* \end{pmatrix} \delta^{I}_-.
\end{align}
This indeed satisfies
\begin{align}
	\epsilon_{IJ} \vert {p}\rangle_a^{I}[{p}\vert_{\dot{a}}^{J} = p_{a\dot{a}},
\end{align}
with $\epsilon_{+-} = -1$ (or $``+" = 1$ and $``-" = 2$).
We therefore define
\begin{align}
	&\vert i\rangle_a = \sqrt{E_i+p_i}\begin{pmatrix} -s_i^* \\ c_i \end{pmatrix},
	\quad
	[i\vert_{\dot{a}} = \sqrt{E_i + p_i}\begin{pmatrix} -s_i & c_i \end{pmatrix},
	\\
	&\vert \eta_i \rangle_a = \sqrt{E_i-p_i}\begin{pmatrix} c_i \\ s_i \end{pmatrix},
	\quad
	[\eta_i\vert_{\dot{a}} = - \sqrt{E_i-p_i}\begin{pmatrix} c_i & s_i^* \end{pmatrix},
\end{align}
which satisfy
\begin{align}
	\langle i \eta_i \rangle = [i \eta_i] = m_i.
\end{align}
The momentum is written as
\begin{align}
	(p_i)_{a\dot{a}} = \vert i\rangle_a [i\vert_{\dot{a}} - \vert\eta_i\rangle_a [\eta_i\vert_{\dot{a}},
\end{align}
where
\begin{align}
	\vert {i}\rangle^{+} = \vert \eta_i \rangle,
	~~
	\vert {i}\rangle^{-} = \vert i\rangle,
	~~
	[{i}\vert^{+} = [i\vert,
	~~
	[{i}\vert^{-} = [\eta_{i}\vert.
\end{align}
The polarization vectors of the gauge bosons are defined in the helicity basis as
\begin{align}
	\epsilon_i^{(+)} = \sqrt{2}\frac{\vert \eta_i\rangle [ i \vert}{m_i},
	\quad
	\epsilon_i^{(-)} = -\sqrt{2}\frac{\vert i\rangle [ \eta_i \vert}{m_i},
	\quad
	\epsilon_i^{(L)} = \frac{\vert i\rangle [i \vert + \vert \eta_i\rangle [\eta_i \vert}{m_i}.
\end{align}
The norms are given by
\begin{align}
	\epsilon^{(+)}_i \cdot \epsilon^{(-)}_i = \epsilon^{(L)}_i \cdot \epsilon^{(L)}_i = -1.
\end{align}
Note that, in a general basis, the polarization vector has two little group indices symmetrized as
\begin{align}
	[\epsilon_i^{IJ}]_{a\dot{a}} = \frac{\sqrt{2}}{m_i}\vert {i}\rangle_a^{\{I} [{i}\vert_{\dot{a}}^{J\}}.
\end{align}
The above three polarization vectors correspond to 
\begin{align}
	\epsilon_i^{++} = \epsilon_i^{(+)},
	\quad
	\epsilon_i^{--} = -\epsilon_i^{(-)},
	\quad
	\epsilon_i^{\{+,-\}} = \epsilon_i^{(L)},
	\label{eq:pol_massive_relation}
\end{align}
where the symmetrization, $\{+,-\}$, is understood to involve the factor $1/\sqrt{2}$ in its normalization.
For a massless particle such as the photon, we can define the limit
\begin{align}
	\lim_{m_i\to 0} \frac{\vert \eta_i \rangle}{m_i} = \vert \xi_i \rangle,
	\quad
	\lim_{m_i\to 0} \frac{\vert \eta_i ]}{m_i} = \vert \xi_i ],
\end{align}
where $\vert \xi_i \rangle$ and $\vert \xi_i ]$ are the arbitrary reference spinors.
With them, the polarization vector is given by
\begin{align}
	\epsilon_i^{(+)} = \sqrt{2}\frac{\vert \xi_i\rangle [ i \vert}{\langle i \xi_i\rangle},
	\quad
	\epsilon_i^{(-)} = -\sqrt{2}\frac{\vert i\rangle [ \xi_i \vert}{[i\xi_i]}.
\end{align}
Note that the polarization vectors satisfy the completeness relation,
\begin{align}
	\sum_{\lambda=\pm,L} [\epsilon_{i}^{(\lambda)}]_{a\dot{a}} [\epsilon_{i}^{(-\lambda)}]_{b\dot{b}}
	= -2\epsilon_{ab}\epsilon_{\dot{a}\dot{b}} + \frac{[p_i]_{a\dot{a}}[p_i]_{b\dot{b}}}{m_i^2},
	\label{eq:pol_comp_massive_spinor}
\end{align}
for the massive case, and
\begin{align}
	\sum_{\lambda=\pm} [\epsilon_{i}^{(\lambda)}]_{a\dot{a}} [\epsilon_{i}^{(-\lambda)}]_{b\dot{b}}
	= -2\epsilon_{ab}\epsilon_{\dot{a}\dot{b}} 
	+ \frac{[p_i]_{a\dot{a}}[\bar{p}_i]_{b\dot{b}} + [\bar{p}_i]_{a\dot{a}}[{p}_i]_{b\dot{b}}}{p_i\cdot\bar{p}_i},
	\quad
	\bar{p}_i = \vert \xi_i\rangle [\xi_i\vert,
	\label{eq:pol_comp_massless_spinor}
\end{align}
for the massless case, respectively.

Going back to the vector notation, these are given by
\begin{align}
	\sum_{\lambda=\pm,L} [\epsilon_{i}^{(\lambda)}]_{\mu} [\epsilon_{i}^{(-\lambda)}]_{\nu}
	= -\eta_{\mu\nu} + \frac{[p_i]_{\mu}[p_i]_{\nu}}{m_i^2},
	\label{eq:pol_comp_massive_vector}
\end{align}
for the massive case, and
\begin{align}
	\sum_{\lambda=\pm} [\epsilon_{i}^{(\lambda)}]_{\mu} [\epsilon_{i}^{(-\lambda)}]_{\nu}
	= -\eta_{\mu\nu}
	+ \frac{[p_i]_{\mu}[\bar{p}_i]_{\nu} + [\bar{p}_i]_{\mu}[{p}_i]_{\nu}}{p_i\cdot\bar{p}_i},
	\label{eq:pol_comp_massless_vector}
\end{align}
for the massless case.

Polarization vectors and fermion wavefunctions are eigenstates of spin operator $J_k^0$ where spin-axis is aligned along momentum $k^{\mu}$, and they are related by spin-raising and lowering operator $J^{\pm}_k$. These operators are the three generators of the little group SU(2) for real momentum,
\begin{align}
 J_k^0 &=  \frac{1}{2} \left(-\lvert k \rangle_\alpha \dfrac{\partial}{\partial \lvert k \rangle_\alpha} + \lvert \eta_k \rangle_\alpha \dfrac{\partial}{\partial \lvert \eta_k \rangle_\alpha} + [ k \lvert_{\dot{\alpha}} \dfrac{\partial}{\partial [ k \lvert_{\dot{\alpha}} } - [  \eta_k \lvert_{\dot{\alpha}} \dfrac{\partial}{\partial [ \eta_k \lvert_{\dot{\alpha}}}    \right), 
 	\\
     J_k^- &= -\left(  \lvert k \rangle_\alpha \dfrac{\partial}{\partial \lvert \eta_k \rangle_\alpha} + [ \eta_k \lvert_{\dot{\alpha}} \dfrac{\partial}{\partial [ k\lvert_{\dot{\alpha}}}    \right), 
     \quad
      J_k^+ = \left(  \lvert \eta_k \rangle_\alpha \dfrac{\partial}{\partial \lvert k \rangle_\alpha} + [ k \lvert_{\dot{\alpha}} \dfrac{\partial}{\partial [ \eta_k\lvert_{\dot{\alpha}}}  \right). 
\end{align}
The commutation relations among these differential operators are
\begin{align}
    [J_k^0,J_k^{\pm}] = \pm J_k^{\pm}, \qquad  [J_k^+,J_k^-] = -2 J_k^{0}
\end{align}
and so we observe that $J^{\pm}_k$ serve as the spin raising and lowering operators for the eigenstate of $J_k^0$. 
We can apply spin operators to the polarization tensors as
\begin{align}
	J_k^0 \epsilon_k^{(\pm)} &= \pm\epsilon_k^{(\pm)}, \quad
	J_k^0 \epsilon_k^{(L)} = 0, \quad
	J^{\mp}_k \epsilon_k^{(\pm)} = -\frac{1}{\sqrt{2}}\epsilon_k^{(L)},
	\quad 
	J_k^{\pm}\epsilon_k^{(L)} =\sqrt{2} \epsilon_k^{(\pm)}, 
	\quad 
	J^{\pm}_k p_k = 0.
\end{align}
%%%%%%%%%%%%%%%%%%%%%%%%%%%%%%%%%%%%%%%
\section{On-shell construction of electroweak theory in vector representation}
\label{app:EW_vector}
%%%%%%%%%%%%%%%%%%%%%%%%%%%%%%%%%%%%%%%

For readers' convenience, in this appendix, 
we express the on-shell computation of the EW gauge boson scatterings 
in Sec.~\ref{sec:EW} in the vector basis.

%%%%%%%%%%%%%%%%%%%%%%%%%%%%%%%%%%%%%%%
\subsection{Three-point amplitude}
%%%%%%%%%%%%%%%%%%%%%%%%%%%%%%%%%%%%%%% 

The three-point amplitudes  involving only the gauge bosons, defined in Sec.~\ref{subsec:three-point_EW}, 
are expressed in the vector basis as
\begin{align}
	A_{WWZ}^{(\lambda_1\lambda_2\lambda_3)} &= 
	g_{WWZ} \epsilon_{\alpha}^{(\lambda_1)}(p_1) \epsilon_{\beta}^{(\lambda_2)}(p_2)
	V^{\alpha\beta\mu}(p_1,p_2,p_3) \epsilon_{\mu}^{(\lambda_3)}(p_3),
	\\
	A_{WW\gamma}^{(\lambda_1\lambda_2\lambda_3)} &= 
	g_{WW\gamma} \epsilon_{\alpha}^{(\lambda_1)}(p_1) \epsilon_{\beta}^{(\lambda_2)}(p_2)
	V^{\alpha\beta\mu}(p_1,p_2,p_3) \epsilon_{\mu}^{(\lambda_3)}(p_3),
\end{align}
where $p_3 = -p_1 - p_2$ and
\begin{align}
	V_{\alpha\beta\mu}(p_1,p_2,p_3) = \eta_{\alpha\beta}(p_1 - p_2)_\mu + \eta_{\beta\mu}(p_2 -p_3)_\alpha
	+ \eta_{\mu\alpha}(p_3 - p_1)_\beta.
\end{align}
The Ward identity~\eqref{eq:VVV_Ward} is given by
\begin{align}
	&\left.A_{WWZ}\right\vert_{\epsilon_k \to p_k}
	= -g_{WWZ}(m_i^2 -m_j^2) \epsilon_i \cdot \epsilon_j,
	\quad
	\left.A_{WW\gamma}\right\vert_{\epsilon_3 \to p_3}
	= 0.
	\label{eq:VVV_Ward_vector}
\end{align}
%%

%%%%%%%%%%%%%%%%%%%%%%%%%%%%%%%%%%%%%%%
\subsection{$WWWW$ amplitude}
%%%%%%%%%%%%%%%%%%%%%%%%%%%%%%%%%%%%%%% 

By using the completeness relations~\eqref{eq:pol_comp_massive_vector} and~\eqref{eq:pol_comp_massless_vector}, 
Eq.~\eqref{eq:pole_WWWW} is given by
\begin{align}
	A_{WWWW}^{(\lambda_1 \lambda_2 \lambda_3\lambda_4)}
	&= -\sum_{I = Z, \gamma} \sum_{i=2,4}\frac{g_{WWI}^2}{p_{1i}^2-m_I^2}
	\frac{1}{{z^+_{1i} - z^-_{1i}}}
	\nonumber \\
	&\times
	\epsilon_{1}^\alpha \epsilon_i^\beta 
	\left[
	{z^+_{1i}} 
	\left[{V_{\alpha\beta}}^\mu(\hat{p}_1, \hat{p}_i, -\hat{p}_{1i})
	\times 
	V_{\gamma\delta \mu}(\hat{p}_3, \hat{p}_j, -\hat{p}_{3j})
	\right]_{z = z^-_{1i}}
	- (z_{1i}^+ \leftrightarrow z_{1i}^-)
	\right]
	\epsilon_3^\gamma \epsilon_j^\delta.
\end{align}
The product of $V$ contains up to the quadratic terms in $z$,
and we obtain
\begin{align}
	A_{WWWW}^{(\lambda_1 \lambda_2 \lambda_3\lambda_4)}
	=& -\sum_{I = Z, \gamma} \sum_{i=2,4}\frac{g_{WWI}^2}{p_{1i}^2-m_I^2}
	\epsilon_{1}^\alpha \epsilon_i^\beta 
	{V_{\alpha\beta}}^\mu({p}_1, {p}_i, -{p}_{1i})
	\times 
	V_{\gamma\delta \mu}({p}_3, {p}_j, {p}_{1i})
	\epsilon_3^\gamma \epsilon_j^\delta
	\nonumber \\
	&
	+\sum_{i=2,4} \frac{g^2}{2q_1 \cdot q_i}
	\epsilon_{1}^\alpha \epsilon_i^\beta 
	{V_{\alpha\beta}}^\mu(q_1, q_i, -q_{1i})
	\times 
	V_{\gamma\delta \mu}(q_3, q_j, -q_{3j})
	\epsilon_3^\gamma \epsilon_j^\delta,
\end{align}
which is equivalent to Eq.~\eqref{eq:4W_bfr_simplification},
where the first term is from the $z^0$-term while the second term is from the $z^2$-term.
To simplify the second term, we remind ourselves that we take the particle~1 as transverse, $\lambda_1 = \pm$,
and hence $\epsilon_1^\alpha = {q_1^\alpha}/{c_1 m_1}$.
We then obtain
\begin{align}
	\sum_{i=2,4} \frac{1}{2q_1 \cdot q_i}
	\epsilon_{1}^\alpha \epsilon_i^\beta 
	{V_{\alpha\beta}}^\mu(q_1, q_i, -q_{1i})
	\times 
	V_{\gamma\delta \mu}(q_3, q_j, -q_{3j})
	\epsilon_3^\gamma \epsilon_j^\delta
	= - \epsilon_{1}^\alpha \epsilon_2^\beta 
	\left[2\eta^{\alpha\gamma}\eta^{\beta\delta} - \eta^{\alpha\beta}\eta^{\gamma\delta}
	-\eta^{\alpha\delta}\eta^{\beta\gamma}\right]
	\epsilon_3^\gamma \epsilon_4^\delta.
\end{align}
Therefore, combining all the terms, we obtain
\begin{align}
	A_{WWWW}^{(\lambda_1 \lambda_2 \lambda_3\lambda_4)}
	=& -\sum_{i=2,4}
	\left[\frac{g_{WWZ}^2}{p_{1i}^2-m_Z^2} + \frac{g_{WW\gamma}^2}{p_{1i}^2}\right]
	\epsilon_{1}^\alpha \epsilon_i^\beta 
	{V_{\alpha\beta}}^\mu({p}_1, {p}_i, -{p}_{1i}) 
	V_{\gamma\delta \mu}({p}_3, {p}_j, {p}_{1i})
	\epsilon_3^\gamma \epsilon_j^\delta
	\nonumber \\
	&
	- g^2\epsilon_{1}^\alpha \epsilon_2^\beta 
	\left[2\eta^{\alpha\gamma}\eta^{\beta\delta} - \eta^{\alpha\beta}\eta^{\gamma\delta}
	-\eta^{\alpha\delta}\eta^{\beta\gamma}\right]
	\epsilon_3^\gamma \epsilon_4^\delta,
\end{align}
which is equivalent to Eq.~\eqref{eq:EW_WWWW_final}, expressed in the vector basis.

%%%%%%%%%%%%%%%%%%%%%%%%%%%%%%%%%%%%%%%
\subsection{$WWZZ$ amplitude}
%%%%%%%%%%%%%%%%%%%%%%%%%%%%%%%%%%%%%%% 

We use the completeness relation~\eqref{eq:pol_comp_massive_vector} to write Eq.~\eqref{eq:pole_WWZZ} as
\begin{align}
	A_{WWZZ}^{(\lambda_1 \lambda_2 \lambda_3\lambda_4)}
	&=-g_{WWZ}^2\sum_{i=3,4}\frac{1}{p_{1i}^2-m_W^2}
	\frac{1}{{z^+_{1i} - z^-_{1i}}}
	\nonumber \\
	&\times
	\epsilon_{1}^\alpha \epsilon_i^\gamma 
	\left[
	{z^+_{1i}} 
	\left[V_{\alpha\mu\gamma}(\hat{p}_1, -\hat{p}_{1i}, \hat{p}_i ) 
	\left(\eta^{\mu\nu} - \frac{\hat{p}_{1i}^\mu\hat{p}_{1i}^\nu}{m_W^2}\right)
	V_{\nu\beta\delta}(-\hat{p}_{2j}, \hat{p}_2, \hat{p}_j)
	\right]_{z = z^-_{1i}}
	- (z_{1i}^+ \leftrightarrow z_{1i}^-)
	\right]
	\epsilon_2^\beta \epsilon_j^\delta.
\end{align}
With the help of the Ward identity of the three-point amplitudes, we rewrite it as
\begin{align}
	A_{WWZZ}^{(\lambda_1 \lambda_2 \lambda_3\lambda_4)}
	=& -g_{WWZ}^2\sum_{i=3,4}\frac{1}{p_{1i}^2-m_W^2}
	\epsilon_{1}^\alpha \epsilon_i^\gamma
	\left[V_{\alpha\mu\gamma}({p}_1, -{p}_{1i}, {p}_i )
	\times 
	\left(\eta^{\mu\nu} - \frac{{p}_{1i}^\mu{p}_{1i}^\nu}{m_W^2}\right)
	V_{\nu\beta\delta}(-{p}_{2j}, {p}_2, {p}_j)
	\right]
	\epsilon_2^\beta \epsilon_j^\delta
	\nonumber \\
	&+ \sum_{i=3,4}\frac{g_{WWZ}^2}{2q_1 \cdot q_i}
	\epsilon_{1}^\alpha \epsilon_i^\gamma
	V_{\alpha\mu\gamma}(q_1, -q_{1i}, q_i )
	{V^\mu}_{\beta\delta}(-q_{2j}, q_2, q_j)
	\epsilon_2^\beta \epsilon_j^\delta,
\end{align}
which corresponds to Eq.~\eqref{eq:WWZZ_bfr_simplification}, 
where again the first term comes from $z^0$ while the second term from $z^2$.
In the same way as the $WWWW$ scattering, we can simplify the second term by assuming that $\lambda_1 = \pm$,
and we obtain
\begin{align}
	A_{WWZZ}^{(\lambda_1 \lambda_2 \lambda_3\lambda_4)}
	=& -g_{WWZ}^2\sum_{i=3,4}\frac{1}{p_{1i}^2-m_W^2}
	\epsilon_{1}^\alpha \epsilon_2^\beta 
	\left[V_{\alpha\mu\gamma}({p}_1, -{p}_{1i}, {p}_i )
	\times 
	\left(\eta^{\mu\nu} - \frac{{p}_{1i}^\mu{p}_{1i}^\nu}{m_W^2}\right)
	V_{\nu\beta\delta}(-{p}_{2j}, {p}_2, {p}_j)
	\right]
	\epsilon_i^\gamma \epsilon_j^\delta
	\nonumber \\
	&+g_{WWZ}^2
	\epsilon_{1}^\alpha \epsilon_2^\beta 
	\left[2\eta_{\alpha\beta}\eta_{\gamma\delta} - \eta_{\alpha\gamma}\eta_{\beta\delta}
	-\eta_{\alpha\delta}\eta_{\beta\gamma}\right]
	\epsilon_3^\gamma \epsilon_4^\delta.
\end{align}
This corresponds to Eq.~\eqref{eq:EW_WWZZ_final}, expressed in the vector basis.

%%%%%%%%%%%%%%%%%%%%%%%%%%%%%%%%%%%%%%%
\subsection{$WWtt$ amplitude}
%%%%%%%%%%%%%%%%%%%%%%%%%%%%%%%%%%%%%%% 

For the $WWtt$ amplitude, additional fermions in the spectrum introduce the following three-point interactions
\begin{align}
    A_{tbW}^{(\lambda_1\lambda_2\lambda_3 )} &= g_{tbW}\bar{v}(p_2) \gamma_{\mu} \epsilon^{\mu}(p_3)P_L {u}(p_1), \\
    A_{ttZ}^{(\lambda_1\lambda_2\lambda_3 )} &= \bar{v}(p_2) \gamma_{\mu} \epsilon^{\mu}(p_3)(g_{ttZ}^L P_L+g_{ttZ}^R P_R){u}(p_1),  \\
    A_{tt\gamma}^{(\lambda_1\lambda_2\lambda_3 )} &= g_{tt\gamma}\bar{v}(p_2) \gamma_{\mu} \epsilon^{\mu}(p_3){u}(p_1).
\end{align}
Then we can write Eq.~(\ref{eq:pole_WWtt}) as 
\begin{align}
	A_{WWtt}^{(\lambda_1 \lambda_2 \lambda_3 \lambda_4)}
	&= -\sum_{I=Z,\gamma} \frac{1}{p_{12}^2-m_I^2}
	\frac{1}{{z^+_{12} - z^-_{12}}} \nonumber \\
 &\times \epsilon_{1}^{ \alpha}\epsilon_{2}^{ \beta}\bar{v}_{4} \left( z_{12}^+
 \left[{g_{WWI}V_{\alpha\beta}}^{\mu}(\hat{p}_1, \hat{p}_2, -\hat{p}_{12})
	  \gamma_{\mu} (g_{ttI}^L P_L+g_{ttI}^R P_R) \right]_{z = z^-_{12}}
	- (z_{12}^+ \leftrightarrow z_{12}^-) \right)u_3
	\nonumber \\
	&-\frac{1}{p_{14}^2-m_b^2}
	\frac{1}{{z^+_{14} - z^-_{14}}} \nonumber\\
 &\times \epsilon_{1}^{ \alpha}\epsilon_{2}^{ \beta}\bar{v}_{4} \left( z_{14}^+
 \left[g_{tbW}^2 \gamma_{\alpha}P_L\hat{p}^{\mu}_{14}\gamma_{\mu} \gamma_{\beta} P_L \right]_{z = z^-_{14}}
	- (z_{14}^+ \leftrightarrow z_{14}^-) \right)u_3.
	\label{eq:pole_WWtt_vec}
\end{align}
We notice that each term contains one insertion of momentum in the numerator either from the function ${V_{\alpha\beta}}^{\mu}(\hat{p}_1, \hat{p}_2, -\hat{p}_{12})$ or from internal fermion. These in turn will generate terms linear in $z$ or terms independent in $z$. Summing over all poles, linear terms in $z$ cancels and we find the total amplitude is of the form
\begin{align}
	A_{WWtt}^{(\lambda_1 \lambda_2 \lambda_3 \lambda_4)}
	&= -\sum_{I=Z,\gamma} \frac{1}{p_{12}^2-m_I^2}  \epsilon_{1}^{ \alpha}\epsilon_{2}^{ \beta}\bar{v}_{4} \left( 
{g_{WWI}V_{\alpha\beta}}^{\mu}({p}_1, {p}_2, -{p}_{12})
	  \gamma_{\mu} \left[g_{ttI}^L P_L+g_{ttI}^R P_R \right]
	 \right)u_3
	\nonumber \\
	& \qquad -\frac{1}{p_{14}^2-m_b^2}
 \epsilon_{1}^{ \alpha}\epsilon_{2}^{ \beta}\bar{v}_{4} \left( 
 g_{tbW}^2 \gamma_{\alpha}P_L {p}^{\mu}_{14}\gamma_{\mu}  \gamma_{\beta} P_L 
	 \right)u_3.
	\label{eq:WWtt_vec_1}
\end{align}
This is the vector representation of Eq.~\eqref{eq:WWtt_spin_rep}.
%%%%%%%%%%%%%%%%%%%%%%%%%%%%%%%%%%%%%%%
\section{UV completion of electroweak theory in vector representation}
\label{app:UV_vector}
%%%%%%%%%%%%%%%%%%%%%%%%%%%%%%%%%%%%%%%

Again for readers' convenience, in this appendix, 
we express the on-shell computation of the UV completion of the EW theory in Sec.~\ref{sec:UV} 
in the vector basis.

%%%%%%%%%%%%%%%%%%%%%%%%%%%%%%%%%%%%%%%
\subsection{A neutral scalar: Higgs doublet}
%%%%%%%%%%%%%%%%%%%%%%%%%%%%%%%%%%%%%%%

We begin with including one neutral real scalar field $h$, which we call the Higgs, 
discussed in Sec.~\ref{subsec:UV_doublet}.
The three-point amplitudes are expressed as
\begin{align}
	&A_{WWh}^{(\lambda_1,\lambda_2)} = g_{WWh} \epsilon_1 \cdot \epsilon_2,
	\quad
	A_{ZZ h}^{(\lambda_1,\lambda_2)} = g_{ZZh} \epsilon_1 \cdot \epsilon_2, \quad A_{ffh}^{(\lambda_1\lambda_2)} = g_{ffh} \bar{v}_2 u_1, 
\end{align}
Therefore the amplitudes involving only the gauge bosons are given by
\begin{align}
	\left.A_{WWWW}^{(\lambda_1\lambda_2\lambda_3\lambda_4)}\right\vert_{h}
	&= g_{WWh}^2
	\sum_{i=2,4}
	\frac{(\epsilon_1 \cdot \epsilon_i)(\epsilon_3\cdot \epsilon_j)}{p_{1i}^2 - m_{h}^2},
	\\
	\left.A_{WWZZ}^{(\lambda_1\lambda_2\lambda_3\lambda_4)}\right\vert_{h}
	&= g_{WWh} g_{ZZh}
	\frac{(\epsilon_1 \cdot \epsilon_2)(\epsilon_3\cdot\epsilon_4)}{p_{12}^2 - m_{h}^2},
\end{align}
where $j\neq i$ and $j = 2,4$ in the above sum. These correspond to Eqs.~\eqref{eq:WWWW_higgs_doublet} 
and~\eqref{eq:WWZZ_higgs_doublet}.
The $WWZh$ scattering contains the term only up to linear in $z$ in the numerator at the pole,
and therefore it is given by a simple product of the three-point amplitudes without the momentum shift,
as we have explained in the main text.
Therefore we obtain
\begin{align}
	A_{WWZh}^{(\lambda_1\lambda_2\lambda_3)}
	=& -\frac{g_{WWZ}g_{ZZh}}{p_{12}^2 - m_Z^2}
	\epsilon_1^\alpha\epsilon_2^\beta V_{\alpha\beta\gamma}(p_1,p_2,-p_{12})\epsilon_3^\gamma
	\nonumber \\
	&+ \sum_{i=1,2}
	(-1)^{i}\frac{g_{WWZ}g_{WWh}}{p_{i3}^2 - m_W^2} \epsilon_i^\alpha \epsilon_3^\beta V_{\alpha\mu\beta}
	(p_i,-p_{i3},p_3)\left(\eta^{\mu\nu}-\frac{p_{i3}^\mu p_{i3}^\nu}{m_W^2}\right)\epsilon_{j\nu},
\end{align}
in the vector notation where $j \neq i, 3,4$. This corresponds to Eq.~\eqref{eq:WWZh_higgs_doublet}.
Finally, after using the completeness relation~\eqref{eq:pol_comp_massive_vector}, 
the $WWhh$ amplitude~\eqref{eq:WWhh_doublet_pole} is written as
\begin{align}
	A_{WWhh}^{(\lambda_1 \lambda_2)}
	&=
	-\sum_{i=3,4}
	\epsilon_\mu^{(\lambda_1)}\left[\frac{g_{WWh}^2 m_W^2}{p_{1i}^2-m_W^2}
	\left(\eta^{\mu\nu} - \frac{{p}_{1i}^\mu{p}_{1i}^\nu}{m_W^2}\right)
	+ g_{WWh}^2  \frac{q_{1i}^\mu q_{1i}^\nu}{2q_1 \cdot q_i}\right]\epsilon_\nu^{(\lambda_2)},
\end{align}
which corresponds to Eq.~\eqref{eq:WWhh_bfr_simplificaiton_doublet}.
By taking $\epsilon_1 = q_1/c_1 m_W$, we obtain
\begin{align}
	A_{WWhh}^{(\lambda_1 \lambda_2)}
	&= 
	-\epsilon_\mu^{(\lambda_1)}\left[\sum_{i=3,4}\frac{g_{WWh}^2 m_W^2}{p_{1i}^2-m_W^2}
	\left(\eta^{\mu\nu} - \frac{{p}_{1i}^\mu{p}_{1i}^\nu}{m_W^2}\right)
	+ \frac{g_{WWh}^2}{2}\eta^{\mu\nu}\right]\epsilon_\nu^{(\lambda_2)}.
\end{align}
This corresponds to Eq.~\eqref{eq:WWhh_final_doublet}. For $WWtt$ calculations, in addition to Eq.~\eqref{eq:WWtt_vec_1}, we must add the Higgs contribution which is of the form
\begin{align}
A_{WWtt}^{(\lambda_1\lambda_2\lambda_3\lambda_4)} = g_{tth}g_{WWh} \frac{(\epsilon_1 \cdot \epsilon_2) \bar{v}_4 u_3}{p_{12}^2 -m_h^2}
\end{align}
This corresponds to Eq.~\eqref{eq:WWtt_higgs_cont}.

%%%%%%%%%%%%%%%%%%%%%%%%%%%%%%%%%%%%%%%
\subsection{Neutral and doubly charged scalar: Higgs triplet}
%%%%%%%%%%%%%%%%%%%%%%%%%%%%%%%%%%%%%%%

We next express the Higgs triplet case discussed in Sec.~\ref{subsec:UV_triplet} in the vector notation.
The three-point amplitudes are given by
\begin{align}
	&A_{WWh^0}^{(\lambda_1,\lambda_2)} = g_{WWh}^{(0)} \epsilon_1 \cdot \epsilon_2,
	\quad
	A_{ZZ h^0}^{(\lambda_1,\lambda_2)} = g_{ZZh}^{(0)} \epsilon_1 \cdot \epsilon_2,
	\quad
	A_{WWh^{\mp\mp}}^{(\lambda_1,\lambda_2)} = g_{WWh}^{(2)} \epsilon_1 \cdot \epsilon_2,
\end{align}
for those involving two gauge bosons, 
and
\begin{align}
	&A_{h^{++} h^{--} Z}^{(\lambda_3)} = g_{hhZ}^{(2)} (p_1 - p_2)\cdot \epsilon_3,
	\quad
	A_{h^{++} h^{--} \gamma}^{(\lambda_3)} = g_{hh\gamma}^{(2)} (p_1 - p_2)\cdot \epsilon_3,
\end{align}
for those involving one gauge boson, respectively.
Eqs.~\eqref{eq:WWWW_higgs_triplet} and~\eqref{eq:WWZZ_higgs_triplet} are then expressed in the vector basis as
\begin{align}
	\left.A_{WWWW}^{(\lambda_1\lambda_2\lambda_3\lambda_4)}\right\vert_{h}
	&= \left(g_{WWh}^{(0)}\right)^2
	\left[\sum_{i=2,4}
	\frac{(\epsilon_1 \cdot \epsilon_i)(\epsilon_3\cdot \epsilon_j)}{p_{1i}^2 - m_{0}^2}\right]
	+ \left(g_{WWh}^{(2)}\right)^2
	\frac{(\epsilon_1 \cdot \epsilon_3)(\epsilon_2\cdot \epsilon_4)}{p_{13}^2 - m_{2}^2},
	\\
	\left.A_{WWZZ}^{(\lambda_1\lambda_2\lambda_3\lambda_4)}\right\vert_{h}
	&=
	g_{WWh}^{(0)} g_{ZZh}^{(0)}
	\frac{(\epsilon_1 \cdot \epsilon_2)(\epsilon_3\cdot\epsilon_4)}{p_{12}^2 - m_{0}^2},
\end{align}
where $j\neq 1, 3, i$.
For $W^+ W^+ Z h^{--}$ and $W^+W^+ \gamma h^{--}$, 
by using the completeness relation of the intermediate gauge bosons, we obtain
\begin{align}
	A_{WWZh^{--}}^{(\lambda_1\lambda_2\lambda_3)}
	&= - \frac{2g_{WWh}^{(2)}g_{hhZ}^{(2)}}{p_{12}^2 -m_{2}^2} (\epsilon_1 \cdot \epsilon_2)
	(\epsilon_3 \cdot p_4)
	-\sum_{i=1,2}
	\left[
	\frac{g_{WWZ}g_{WWh}^{(2)}}{p_{i3}^2 - m_W^2} \epsilon_i^\alpha \epsilon_3^\beta V_{\alpha\mu\beta}
	(p_i,-p_{i3},p_3)\left(\eta^{\mu\nu}-\frac{p_{i3}^\mu p_{i3}^\nu}{m_W^2}\right)\epsilon_{j\nu}
	\right],
	\\
	A_{WW\gamma h^{--}}^{(\lambda_1\lambda_2\lambda_3)}
	&= - \frac{2g_{WWh}^{(2)}g_{hh\gamma}^{(2)}}{p_{12}^2 -m_{2}^2} (\epsilon_1 \cdot \epsilon_2)
	(\epsilon_3 \cdot p_4)
	-\sum_{i=1,2}
	\left[
	\frac{g_{WW\gamma}g_{WWh}^{(2)}}{p_{i3}^2 - m_W^2} \epsilon_i^\alpha \epsilon_3^\beta V_{\alpha\mu\beta}
	(p_i,-p_{i3},p_3)\left(\eta^{\mu\nu}-\frac{p_{i3}^\mu p_{i3}^\nu}{m_W^2}\right)\epsilon_{j\nu}
	\right],
\end{align}
in the vector notation where $j \neq 3,4, i$.
These correspond to Eqs.~\eqref{eq:WWZh_triplet} and~\eqref{eq:WWgammah_triplet}.
In particular, the Ward identity of the $W^+W^+ \gamma h^{--}$ amplitude requires
\begin{align}
	g_{hh\gamma}^{(2)} = -2g_{WW\gamma}.
\end{align}
Finally, the $W^+W^- h_{a}^{++}h_b^{--}$ amplitude is expressed as
\begin{align}
	A_{WWh^{++}h^{--}}^{(\lambda_1\lambda_2)}
	=& -\frac{g_{WWZ}g_{hhZ}^{(2)}}{p_{12}^2-m_Z^2}
	\frac{1}{z_{12}^+ - z_{12}^-}
	\epsilon_1^\alpha\epsilon_2^\beta\left[z_{12}^+\left[
	V_{\alpha\beta\mu}(\hat{p}_1,\hat{p}_2,-\hat{p}_{12}) (\hat{p}_3^\mu-\hat{p}_4^\mu)
	\right]_{z_{12}^-}
	- (z_{12}^+ \leftrightarrow z_{12}^-)\right]
	\nonumber \\
	&+\frac{2g_{WW\gamma}^2}{p_{12}^2}
	\frac{1}{z_{12}^+ - z_{12}^-}
	\epsilon_1^\alpha\epsilon_2^\beta\left[z_{12}^+\left[
	V_{\alpha\beta\mu}(\hat{p}_1,\hat{p}_2,-\hat{p}_{12}) (\hat{p}_3^\mu-\hat{p}_4^\mu)
	\right]_{z_{12}^-}
	- (z_{12}^+ \leftrightarrow z_{12}^-)\right]
	\nonumber \\
	&+ \frac{\left(g_{WWh}^{(2)}\right)^2}{p_{14}^2 - m_W^2}
	\frac{1}{z_{14}^+ - z_{14}^-}
	\epsilon_1^\alpha\epsilon_2^\beta\left[
	z_{12}^+\left[-\eta_{\alpha\beta} + \frac{\hat{p}_{14\alpha}\hat{p}_{14\beta}}{m_W^2}
	\right]_{z_{12}^-}
	- (z_{12}^+ \leftrightarrow z_{12}^-)
	\right].
\end{align}
Each term contain $z^0$ and $z^2$ terms, and by focusing on the $z^2$ term
and taking $q_1 = c_1 m_W \epsilon_1$, we obtain
\begin{align}
	\left[A_{WWh^{++}h^{--}}^{(\lambda_1\lambda_2)}\right]_\mathrm{contact}
	&= \frac{1}{2c_1 m_W^3}
	\left[
	2\left(g_{WWZ}g_{hhZ}^{(2)}  + g_{WW\gamma} g_{hh\gamma}^{(2)}\right)m_W^2 \epsilon_2 \cdot (q_3 - q_4)
	+ \left(g_{WWh}^{(2)}\right)^2 \epsilon_2 \cdot q_3
	\right],
\end{align}
which corresponds to Eq.~\eqref{eq:WWhh_triplet_bfr_cond}.
Therefore, the final result does not depend on these variables only if
\begin{align}
	0 &= 4\left(g_{WWZ}g_{hhZ}^{(2)} + g_{WW\gamma} g_{hh\gamma}^{(2)}\right)m_W^2
	+ \left(g_{WWh}^{(2)}\right)^2,
\end{align}
and with this condition we obtain the contact term as
\begin{align}
	\left[A_{WWh^{++}h^{--}}^{(\lambda_1\lambda_2)}\right]_\mathrm{contact}
	&= -\frac{\left(g_{WWh}^{(2)}\right)^2}{4m_W^2}\epsilon_1 \cdot \epsilon_2.
\end{align}
This corresponds to Eq.~\eqref{eq:WWhh_triplet_contact}.

%%%%%%%%%%%%%%%%%%%%%%%%%%%%%%%%%%%%%%%
%\small
\bibliographystyle{utphys}
\bibliography{ref}
%%%%%%%%%%%%%%%%%%%%%%%%%%%%%%%%%%%%%%%
  
%%%%%%%%%%%%%%%%%%%%%%%%%%%%%%%%%%%%%%%
\end{document}